\documentclass[sigconf]{acmart}

\usepackage{booktabs} 
\usepackage{multirow}
\usepackage{makecell}
\usepackage{xifthen}
\usepackage[nolist,nohyperlinks]{acronym}
\usepackage[ruled,vlined]{algorithm2e}
\usepackage[capitalise]{cleveref}
\usepackage{siunitx}
\usepackage{dsfont}
\usepackage[caption = false]{subfig}

\begin{acronym}
\acro{ICWS}{improved consistent weighted sampling}
\acro{CCWS}{canonical consistent weighted sampling}
\acro{PCWS}{practical consistent weighted sampling}
\acro{I2CWS}[I$^{2}$CWS]{improved \acs*{ICWS}}
\acro{TF-IDF}{term frequency--inverse document frequency}
\acro{PRNG}{pseudorandom number generator}
\acro{MSE}{mean squared error}
\end{acronym}

\newcommand*{\symDefine}[2]{\newcommand{{#1}}{{#2}}}

\symDefine{\symSetA}{A}
\symDefine{\symBBit}{b}
\symDefine{\symSetB}{B}
\symDefine{\symNumExamples}{c}
\symDefine{\symUniverse}{D}
\symDefine{\symUniverseElement}{d}
\symDefine{\symHashValue}{h}
\symDefine{\symHarmonic}{H}
\symDefine{\symHashValueMax}{\symHashValue_\text{max}}
\symDefine{\symHashIndex}{i}
\symDefine{\symSomeIndex}{j}
\symDefine{\symInsertionCounter}{j}
\symDefine{\symJaccard}{J}
\symDefine{\symJaccardEstimator}{\hat{\symJaccard}}
\symDefine{\symDiscreteIndex}{k}
\symDefine{\symWeightIndex}{l}
\symDefine{\symWeightIndexMinus}{p}
\symDefine{\symWeightIndexMinusPrime}{\symWeightIndexMinus'}
\symDefine{\symWeightIndexPlus}{q}
\symDefine{\symWeightIndexPlusPrime}{\symWeightIndexPlus'}
\symDefine{\symWeightIndexCenter}{r}
\symDefine{\symWeightIndexMax}{L}
\symDefine{\symHashSize}{m}
\symDefine{\symInputSize}{n}
\symDefine{\symBigO}{\mathcal{O}}
\symDefine{\symLandauLow}{\Omega}
\symDefine{\symPoisson}{P}
\symDefine{\symIndexSet}{Q}
\symDefine{\symRNG}{R}
\symDefine{\symSetS}{S}
\symDefine{\symSparsity}{s}
\symDefine{\symTestCaseIndex}{u}
\symDefine{\symDiscreteValue}{v}
\symDefine{\symDiscreteSet}{V}
\symDefine{\symWeight}{w}
\symDefine{\symDiscreteWeight}{\tilde{\symWeight}}
\symDefine{\symPoint}{x}
\symDefine{\symRandomVariate}{y}
\symDefine{\symZScore}{z}

\symDefine{\symPropConstant}{\alpha}
\symDefine{\symComplexityFunc}{\varphi}
\symDefine{\symExpRate}{\lambda}
\symDefine{\symEmpiricalMSE}{\widehat{\text{MSE}}}

\newcommand\symIndicator{\operatorname{\mathds{1}}}
\newcommand\symVariance{\operatorname{Var}}
\newcommand\symExpectation{\operatorname{\mathbb{E}}}
\newcommand\symProbability{\operatorname{\mathds{P}}}

\newcommand*{\numsci}[1]{\num[scientific-notation = true, output-exponent-marker = \ensuremath{\mathrm{E}}]{#1}}

\SetKwComment{Com}{}{}
\SetKwProg{Class}{class}{}{end class}
\SetKwBlock{Fields}{fields}{end fields}
\SetKwFunction{PoissonProcess}{PoissonProcess}
\SetKwFunction{NextPoint}{Next}
\SetKwFunction{Split}{Split}
\SetKwFunction{Splittable}{Splittable}
\SetKwFunction{PartiallyRelevant}{PartiallyRelevant}
\SetKwFunction{FullyRelevant}{FullyRelevant}
\SetKwProg{Function}{}{}{}
\SetKw{Break}{break}
\SetKw{New}{new}
\SetKw{And}{and} 
\SetKw{Not}{not} 
\SetArgSty{textnormal}
\SetFuncSty{textnormal}

\SetCommentSty{mycommfont}

\newcommand{\myAlg}[2][]{
\ifthenelse{\isempty{#1}}%
    {\begin{figure}}
    {\begin{figure}[#1]}
\begingroup 
\csname @twocolumnfalse\endcsname
\noindent
\resizebox{\columnwidth}{!}{%
\begin{minipage}{1.33\columnwidth}
\begin{algorithm}[H]
\DontPrintSemicolon
{#2}
\end{algorithm}
\end{minipage}%
}
\endgroup
\end{figure}
}

\renewcommand\footnotetextcopyrightpermission[1]{} 
\setcopyright{none}
\author{Otmar Ertl}
\affiliation{
  \city{Linz}
  \country{Austria}
}
\email{otmar.ertl@gmail.com}

\begin{document}

\sloppy
\title{BagMinHash -- Minwise Hashing Algorithm for Weighted Sets}

\begin{abstract}
Minwise hashing has become a standard tool to calculate signatures which allow direct estimation of Jaccard similarities. While very efficient algorithms already exist for the unweighted case, the calculation of signatures for weighted sets is still a time consuming task. BagMinHash is a new algorithm that can be orders of magnitude faster than current state of the art without any particular restrictions or assumptions on 
weights or data dimensionality. Applied to the special case of unweighted sets, it represents the first efficient algorithm producing independent signature components. A series of tests finally verifies the new algorithm and also reveals limitations of other approaches published in the recent past.
\end{abstract}

\maketitle

\section{Introduction}
Increasing amounts of data require efficient algorithms to keep processing times reasonably small. Sketching algorithms can be used to trade accuracy for less computational costs. In case of tasks such as near-duplicate detection, classification, or clustering, 
which are based on proximity computations between objects, 
locality-sensitive hashing enabled large-scale applications by reducing the dimensionality of high-dimensional data \cite{Leskovec2014}. 

Various distance metrics can be used to model similarity. If objects can be described as sets of features, a very popular metric is the Jaccard distance $1 - \symJaccard$ where the Jaccard similarity $\symJaccard$ of sets $\symSetA$ and  $\symSetB$ is defined as 
\begin{equation}
\label{equ:jaccard_index}
\symJaccard
= 
\frac{
|\symSetA\cap \symSetB|
}{
|\symSetA\cup \symSetB|
}.
\end{equation}
Minwise hashing was the first algorithm to calculate hash signatures (also known as fingerprints) of sets which can be directly used for Jaccard similarity estimation \cite{Broder1997}. It maps a set $\symSetS$ to an $\symHashSize$-dimensional vector $\symHashValue_\symSetS = (\symHashValue_{\symSetS 1},\symHashValue_{\symSetS 2},\ldots,\symHashValue_{\symSetS\symHashSize})$ with statistically independent components. Ignoring potential hash collisions, the probability that same components of two different signatures $\symHashValue_\symSetA$ and $\symHashValue_\symSetB$ for sets $\symSetA$ and $\symSetB$ have identical value is equal to the Jaccard similarity
\begin{equation}
\label{equ:signature}
\symProbability(\symHashValue_{\symSetA\symHashIndex}
=
\symHashValue_{\symSetB\symHashIndex})
=
\symJaccard,
\qquad
\forall\symHashIndex\in\lbrace 1,2,\ldots, \symHashSize\rbrace.
\end{equation}
This property allows unbiased estimation of $\symJaccard$ given the fraction of matching components
\begin{equation}
\label{equ:jaccard_estimator}
\symJaccardEstimator
(\symHashValue_\symSetA,\symHashValue_\symSetB)
=
\frac{1}{\symHashSize}
\sum_{\symHashIndex=1}^\symHashSize \symIndicator(
\symHashValue_{\symSetA\symHashIndex}
=
\symHashValue_{\symSetB\symHashIndex})
\end{equation}
where $\symIndicator$ denotes the indicator function. The variance of this estimator is
\begin{equation}
\label{equ:jaccard_variance}
\symVariance(\symJaccardEstimator
(\symHashValue_\symSetA,\symHashValue_\symSetB)
)
=
\symJaccard (1-\symJaccard) / \symHashSize,
\end{equation}
because the number of equal components is binomially distributed with success probability $\symJaccard$.

The calculation of signatures of size $\symHashSize$ using minwise hashing is expensive, because the computation time scales like $\symBigO(\symHashSize\symInputSize)$ with $\symInputSize$ denoting the cardinality of the input set. Typical values for $\symHashSize$ lie in the range from \num{100} to \num{10000} and $\symInputSize$ can be in the millions \cite{Raff2017}. 

\subsection{Advanced Minwise Hashing Techniques}
\label{sec:advanced_techniques}
One permutation hashing is able to reduce the computational costs significantly and has optimal time complexity $\symBigO(\symHashSize+\symInputSize)$ \cite{Li2012}. However, this approach leads to biased Jaccard estimates for small $\symInputSize$ compared to $\symHashSize$ . This problem can be remedied by introducing a postprocessing step called densification \cite{Shrivastava2017}. Unfortunately, this approach increases the variance of the estimator compared to \eqref{equ:jaccard_variance}, because it breaks the statistical independence of individual signature components leading to correlated occurrences of matching components.

Recent developments such as fast similarity sketching \cite{Dahlgaard2017} with time complexity $\symBigO(\symHashSize \log \symHashSize +\symInputSize)$ or the SuperMinHash algorithm \cite{Ertl2017} with time complexity $\symBigO(\symHashSize \log^2 \symHashSize +\symInputSize)$ also lead to signature component matches that are statistically dependent. But their correlation is negative, which even reduces the variance and thus improves the estimation error for small $\symInputSize$. 
However, all these algorithms cannot replace the original approach, if independence of components is important. 

An orthogonal technique that can be used with any minwise hashing approach is $\symBBit$-bit hashing \cite{Li2010, Li2011}. It maps signature components uniformly to $\symBBit$-bit integer values. The higher probability of matching signature components by chance introduces an estimation bias which is compensated by the estimator
\begin{equation*}
\symJaccardEstimator^{\symBBit}(\symHashValue^{\symBBit}_\symSetA,\symHashValue^{\symBBit}_\symSetB)
=
\frac{
\symJaccardEstimator(\symHashValue^{\symBBit}_\symSetA,\symHashValue^{\symBBit}_\symSetB)
-
2^{-\symBBit}
}
{
1-
2^{-\symBBit}
}
=
\frac{
\left(
\frac{1}{\symHashSize}
\sum_{\symHashIndex=1}^\symHashSize \symIndicator(
\symHashValue^{\symBBit}_{\symSetA\symHashIndex}
=
\symHashValue^{\symBBit}_{\symSetB\symHashIndex})
\right)
-
2^{-\symBBit}
}
{
1-
2^{-\symBBit}
}.
\end{equation*}
Here $\symHashValue^{\symBBit}_\symSetA,\symHashValue^{\symBBit}_\symSetB\in\lbrace 0,1,\ldots,2^\symBBit-1\rbrace^\symHashSize$ denote the signatures after reducing all components to $\symBBit$ bits. The corresponding loss of information increases the variance
\begin{equation*}
\symVariance(\symJaccardEstimator^{\symBBit}(\symHashValue^{\symBBit}_\symSetA,\symHashValue^{\symBBit}_\symSetB)
)
=
\symVariance(\symJaccardEstimator(\symHashValue_\symSetA,\symHashValue_\symSetB))
+
\frac{
1-\symJaccard
}{
(2^\symBBit-1)
\symHashSize}
\end{equation*}
where $\symVariance(\symJaccardEstimator(\symHashValue_\symSetA,\symHashValue_\symSetB))$ denotes the variance of the signature components before reduction to $\symBBit$-bit values, which is given by \eqref{equ:jaccard_variance} in case of original minwise hashing. Hence, to get the same estimation error, the signature size $\symHashSize$ needs to be increased. Nevertheless, if the Jaccard similarity $\symJaccard$ is not too small, the memory footprint of the whole signature can be significantly reduced. However, this comes with higher computational costs for calculating the additional signature components. Therefore, a fast hashing scheme is even more important for $\symBBit$-bit hashing.

\subsection{Weighted Minwise Hashing}
Sometimes it is more convenient to represent objects as bags also known as multisets, where each element is associated with a nonnegative weight. For example, 
words or shingles in text documents are often weighted according to a measure called \ac{TF-IDF} \cite{Leskovec2014}. 
A similarity measure for bags is the weighted Jaccard similarity which generalizes \eqref{equ:jaccard_index}. If two bags are described by weight functions $\symWeight_\symSetA$ and $\symWeight_\symSetB$, respectively, which return a nonnegative weight for each element of a universe $\symUniverse$, the weighted Jaccard similarity is given by
\begin{equation}
\label{equ:generalized_jaccard}
\symJaccard 
= 
\frac{
\sum_{\symUniverseElement\in\symUniverse} \min(\symWeight_\symSetA(\symUniverseElement), \symWeight_\symSetB(\symUniverseElement))
}{
\sum_{\symUniverseElement\in\symUniverse} \max(\symWeight_\symSetA(\symUniverseElement), \symWeight_\symSetB(\symUniverseElement))
}.
\end{equation}
If all weights are binary with values either zero or one, and we define sets $\symSetA$ and $\symSetB$ as
$\symSetA:=\lbrace \symUniverseElement\in\symUniverse:  \symWeight_\symSetA(\symUniverseElement) > 0 \rbrace$ and $\symSetB:=\lbrace \symUniverseElement\in\symUniverse: \symWeight_\symSetB(\symUniverseElement) > 0 \rbrace$, respectively, \eqref{equ:generalized_jaccard} will simplify to \eqref{equ:jaccard_index}.

Finding signatures for the weighted case satisfying \eqref{equ:signature} is more challenging. For integral weights it is possible to add a corresponding number of replications and to apply conventional minwise hashing \cite{Haveliwala2000}. This is not efficient, because elements with large weights blow up the effective data size and hence also the computation time. Some optimizations to reduce the number of replications and to apply unweighted minwise hashing  to real weights are described in \cite{Gollapudi2006} and \cite{Haeupler2014}.

The first algorithm inherently designed for real weights is based on rejection sampling \cite{Kleinberg2002, Charikar2002}. 
A faster approach is consistent weighted hashing with amortized time complexity $\symBigO(\symHashSize \symInputSize)$ \cite{Manasse2010}. Here $\symInputSize=\left|\lbrace\symUniverseElement\in\symUniverse:\symWeight(\symUniverseElement) > 0\rbrace\right|$ denotes the number of elements having nonzero weight. A further development of this concept called \ac{ICWS} achieves the same time complexity as worst case and has become the state-of-the-art algorithm for weighted minwise hashing \cite{Ioffe2010}. 

Recently, a weighted hashing scheme was presented, which can lead to significant speedups \cite{Shrivastava2016}. The algorithm requires the universe $\symUniverse$ to be fixed and that weight upper bounds $\symWeight_\text{max}(\symUniverseElement)$ exist for all $\symUniverseElement\in\symUniverse$. The time complexity of this algorithm depends on the sparsity $\symSparsity$ of the weighted set and is $\symBigO(\symHashSize\symSparsity^{-1} + \symInputSize)$. The sparsity is defined as $\symSparsity := (
\sum_{\symUniverseElement\in\symUniverse} \symWeight(\symUniverseElement)
)/(
\sum_{\symUniverseElement\in\symUniverse} \symWeight_\text{max}(\symUniverseElement))$. If $\symSparsity^{-1}$ is small compared to the number of elements with nonzero weights $(\symSparsity^{-1}\ll\symInputSize)$, the computation time can be significantly smaller compared to the \ac{ICWS} algorithm. However, general applicability is limited by the required a priori knowledge of sharp upper bounds for $\symWeight_\text{max}(\symUniverseElement)$. Moreover, space requirements scale at least linearly with dimensionality $|\symUniverse|$.

\subsection{Applications}

Weighted minwise hashing is used in many applications such as near duplicate image detection \cite{Chum2008}, duplicate news story detection \cite{Alonso2013}, source code deduplication \cite{Markovtsev2017}, time series indexing \cite{Luo2016}, hierarchical topic extraction \cite{Gollapudi2006}, or malware classification \cite{Raff2017} and detection \cite{Drew2017}.
Unfortunately, it was reported that weighted minwise hashing using the \ac{ICWS} algorithm can be very expensive due to the $\symBigO(\symHashSize\symInputSize)$ time complexity, which sometimes limits its application \cite{Drew2017, Raff2017} or requires massive computing power using GPUs \cite{Markovtsev2017}.

Recent works suggest to use weighted minwise hashing for machine learning too. As example, weighted minwise hashing was successfully applied to train generalized min-max kernel support vector machines \cite{Li2017, Li2017a}. Furthermore, asymmetric locality-sensitive hashing \cite{Shrivastava2015}, which can be realized using weighted minwise hashing, was used for efficient deep learning \cite{Spring2017}. Finally, it could also be applied to random forests that are constructed using the weighted Jaccard index as similarity measure \cite{Sathe2017}.

\section{Preliminaries}
The aforementioned improvements on unweighted minwise hashing \cite{Shrivastava2017,Dahlgaard2017,Ertl2017} 
achieve much better time complexities than $\symBigO(\symHashSize \symInputSize)$ of the original approach \cite{Broder1997}. This was our main motivation to develop an algorithm for the weighted case, which reduces the $\symBigO(\symHashSize \symInputSize)$ complexity of the \ac{ICWS} algorithm in a similar fashion.

Both, original minwise hashing as well as the \ac{ICWS} algorithm calculate for every input element $\symHashSize$ different independent hash values, one for each signature component. Hence, $\symInputSize$ different hash values are computed for every signature component. The smallest of them finally defines the corresponding signature value. If $\symInputSize \gg \symHashSize$, an element is expected to generate at most one hash value small enough to contribute to the signature. Therefore, in many cases, it would be sufficient to determine for each input element only the smallest hash value including its assigned signature component. If this could be achieved in a more direct way, 
without the need to calculate all $\symHashSize$ different hash values first, a more efficient signature computation would be possible. 

\subsection{Discretization Error}
\label{sec:discretization}
The \ac{ICWS} algorithm is exact for any nonnegative real weights. This is a nice property from a theoretical point of view. However, computers always approximate real numbers by values from a finite set. Therefore, we analyze the impact of weight discretization on the Jaccard index calculation.
Let $\symDiscreteSet = \lbrace\symDiscreteValue_0, \symDiscreteValue_1, \ldots, \symDiscreteValue_\symWeightIndexMax\rbrace$ be the set of discrete weight values satisfying $0 = \symDiscreteValue_0 < \symDiscreteValue_1<\ldots <\symDiscreteValue_\symWeightIndexMax$. The weight function $\symWeight(\symUniverseElement)$ can be approximated by 
\begin{equation}
\label{equ:discrete_weight}
\symDiscreteWeight(\symUniverseElement):=
\symDiscreteValue_{
\symDiscreteIndex(\symUniverseElement)}
\qquad
\text{with}\quad
\symDiscreteIndex(\symUniverseElement)
:=
\max(\lbrace\symWeightIndex:\symDiscreteValue_\symWeightIndex \leq \symWeight(\symUniverseElement)\rbrace).
\end{equation}
We assume that the discretization $\symDiscreteSet$ is chosen in such a way that for all occurring weights $\symWeight(\symUniverseElement)$
\begin{equation}
\label{equ:value_discretization}
\symDiscreteWeight(\symUniverseElement)\leq 
\symWeight(\symUniverseElement) 
\leq \symDiscreteWeight(\symUniverseElement)(1+\varepsilon)
\end{equation}
is fulfilled, where $\varepsilon\ll1$ is a small positive constant. If an approximation $\tilde{\symJaccard}$ of the generalized Jaccard index \eqref{equ:generalized_jaccard} is calculated using approximations $\symDiscreteWeight_\symSetA$ and $\symDiscreteWeight_\symSetB$ for weight functions $\symWeight_\symSetA$ and $\symWeight_\symSetB$, respectively, following error bounds can be established
\begin{equation}
\symJaccard
(1-\varepsilon)
\leq
\symJaccard
(1+\varepsilon)^{-1}
\leq
\tilde{\symJaccard}
\leq
\symJaccard
(1+\varepsilon).
\end{equation}
They show that the relative error of $\tilde{\symJaccard}$ is limited by $\varepsilon$. Since $\symJaccard\in[0,1]$, $\varepsilon$ is also a bound for the absolute error. 

Assumption \eqref{equ:value_discretization} is nothing special as it is usually satisfied by floating-point numbers as long as $\symWeight(\symUniverseElement)$ is neither subnormal nor too large. As example, assume $\symDiscreteSet$ represents the set of all nonnegative single-precision floating-point numbers for which $\symWeightIndexMax = \text{0x7F7FFFFF} = \num{2139095039}$.
If all occurring nonzero weights are within the normal range of single-precision floating-point numbers which is roughly from \numsci{1.18e-38} to \numsci{3.40e38}, \eqref{equ:value_discretization} will be satisfied and $\varepsilon$ will correspond to the machine epsilon which is approximately \numsci{1.19e-7}.

It makes sense to compare the discretization error $\symJaccard\varepsilon$ to the error introduced by estimating the weighted Jaccard index from hash signatures using \eqref{equ:jaccard_estimator}. According to \eqref{equ:jaccard_variance} the standard error is $\sqrt{\symJaccard(1-\symJaccard)/\symHashSize}$. The discretization error $\symJaccard\varepsilon$ can be ignored as long as $\varepsilon\ll\sqrt{(1-\symJaccard)/(\symJaccard\symHashSize)}$. For instance, consider an extreme case where $\symJaccard$ is close to $1$ and $\symHashSize$ is huge. As example, $\symJaccard= 1-\num{e-6}$ and $\symHashSize= \num{e6}$, for which the estimation error would be in the order of \num{e-6}. The error is still one magnitude larger than the discretization error made by using single-precision weights and can therefore be neglected. This example shows that single precision will be sufficient for most applications. 

The assumption of discrete weights is important for our new approach.
As weight values are already discrete on computers, this is not really a restriction or a disadavantage compared to existing approaches, because one could simply choose $\symDiscreteSet$ based on the given discretization. However, as demonstrated, a coarser choice will often be sufficient as long as the additional discretization error on the Jaccard index remains small compared to the expected estimation error.

\subsection{Random Numbers}
We assume an ideal hash function whose output can be treated as uniformly distributed and truly random number. Hashing the same input multiple times, but each time enriched with a different value from some predefined integer sequence, allows generating a sequence of independent hash values. Therefore, an ideal hash function can be used as ideal \ac{PRNG} which produces random bit sequences of arbitrary length as function of some given seed value. Hereinafter, we may write random number instead of pseudorandom number and always mean some value produced by a \ac{PRNG}.

\section{BagMinHash Algorithm}

In the following we assume some fixed weight discretization $\symDiscreteSet$ as discussed in \cref{sec:discretization}.
As starting point towards our new algorithm, we first consider the hash signature defined by 
\begin{equation}
\label{equ:def_signature}
\symHashValue_{\symHashIndex}
:=
\min_{(\symUniverseElement,\symWeightIndex)\in\symIndexSet}
\symPoint_{\symUniverseElement\symWeightIndex\symHashIndex}
\end{equation}
with $\symIndexSet:=
\bigcup_{\symUniverseElement\in\symUniverse}\bigcup_{\symWeightIndex=1}^{\symDiscreteIndex(\symUniverseElement)}\lbrace(\symUniverseElement,\symWeightIndex)\rbrace
=
\bigcup_{\symUniverseElement\in\symUniverse}
\lbrace
\symUniverseElement
\rbrace
\times
\lbrace
1,2,\ldots,\symDiscreteIndex(\symUniverseElement)\rbrace
$ and $\symDiscreteIndex(\symUniverseElement)$ denoting the function introduced in \eqref{equ:discrete_weight}.
$\symPoint_{\symUniverseElement\symWeightIndex\symHashIndex}$
are independent exponentially distributed pseudorandom numbers
$\symPoint_{\symUniverseElement\symWeightIndex\symHashIndex}\sim
\text{Exponential}(\symPropConstant(\symDiscreteValue_\symWeightIndex - \symDiscreteValue_{\symWeightIndex-1}))$
with rate parameter $\symPropConstant(\symDiscreteValue_\symWeightIndex - \symDiscreteValue_{\symWeightIndex-1})$ where $\symPropConstant$ is some positive constant. The choice of $\symPropConstant$ is arbitrary, because scaling all $\symPoint_{\symUniverseElement\symWeightIndex\symHashIndex}$ by the same factor preserves equalities between signature components $\symHashValue_{\symHashIndex}$.

The probability, that the $\symHashIndex$-th signature component is equal for two different weighted sets $\symSetA$ and $\symSetB$, is given by
\begin{multline}
\label{equ:propbability_equal}
\symProbability(\symHashValue_{\symSetA\symHashIndex}
=
\symHashValue_{\symSetB\symHashIndex})
=
\symProbability\left(
\min_{(\symUniverseElement,\symWeightIndex)\in\symIndexSet_\symSetA}
\symPoint_{\symUniverseElement\symWeightIndex\symHashIndex}
=
\min_{(\symUniverseElement,\symWeightIndex)\in\symIndexSet_\symSetB}
\symPoint_{\symUniverseElement\symWeightIndex\symHashIndex}
\right)=
\\
\symProbability\left(
\min_{(\symUniverseElement,\symWeightIndex)\in\symIndexSet_\symSetA\cap\symIndexSet_\symSetB}
\symPoint_{\symUniverseElement\symWeightIndex\symHashIndex}
<
\min_{(\symUniverseElement,\symWeightIndex)\in\symIndexSet_\symSetA\bigtriangleup\symIndexSet_\symSetB}
\symPoint_{\symUniverseElement\symWeightIndex\symHashIndex}
\right)
=
\symProbability(
\symRandomVariate_{1\symHashIndex}
<
\symRandomVariate_{2\symHashIndex}).
\end{multline}
Here we used that all random variables $\symPoint_{\symUniverseElement\symWeightIndex\symHashIndex}$ must be different as they are drawn from a continuous distribution. $\symRandomVariate_{1\symHashIndex}$ and $\symRandomVariate_{2\symHashIndex}$ are given by
\begin{equation*}
\symRandomVariate_{1\symHashIndex} := \min_{(\symUniverseElement,\symWeightIndex)\in\symIndexSet_\symSetA\cap\symIndexSet_\symSetB}
\symPoint_{\symUniverseElement\symWeightIndex\symHashIndex},
\quad
\symRandomVariate_{2\symHashIndex} := \min_{(\symUniverseElement,\symWeightIndex)\in\symIndexSet_\symSetA\bigtriangleup\symIndexSet_\symSetB}
\symPoint_{\symUniverseElement\symWeightIndex\symHashIndex}.
\end{equation*}
They are both the minimum of independent exponentially distributed random variables. Consequently, $\symRandomVariate_{1\symHashIndex}$ and $\symRandomVariate_{2\symHashIndex}$ are also independent and exponentially distributed. The corresponding rate parameters $\symExpRate_1$ and $\symExpRate_2$ can be calculated as the sum of all contributing rates \cite{Mitzenmacher2005} 
\begin{align*}
\symExpRate_1 
&= 
\sum_{(\symUniverseElement,\symWeightIndex)\in\symIndexSet_\symSetA\cap\symIndexSet_\symSetB}
\symPropConstant(\symDiscreteValue_\symWeightIndex - \symDiscreteValue_{\symWeightIndex-1})
=
\symPropConstant
\sum_{\symUniverseElement\in\symUniverse}
\symDiscreteValue_{
\min(
\symDiscreteIndex_\symSetA(\symUniverseElement),
\symDiscreteIndex_\symSetB(\symUniverseElement)
)}\\
&=\symPropConstant
\sum_{\symUniverseElement\in\symUniverse}
\min(
\symDiscreteValue_{
\symDiscreteIndex_\symSetA(\symUniverseElement)},
\symDiscreteValue_{
\symDiscreteIndex_\symSetB(\symUniverseElement)}
)
=
\symPropConstant
\sum_{\symUniverseElement\in\symUniverse}
\min(
\symDiscreteWeight_\symSetA(\symUniverseElement),
\symDiscreteWeight_\symSetB(\symUniverseElement)
)
\end{align*}
and similarly
\begin{equation*}
\symExpRate_2
=
\symPropConstant
\sum_{\symUniverseElement\in\symUniverse}
\max(
\symDiscreteWeight_\symSetA(\symUniverseElement),
\symDiscreteWeight_\symSetB(\symUniverseElement)
)
-
\min(
\symDiscreteWeight_\symSetA(\symUniverseElement),
\symDiscreteWeight_\symSetB(\symUniverseElement)
)
.
\end{equation*}
Among independent exponentially distributed random numbers, the probability, that one is the smallest, is proportional to the rate parameter of its exponential distribution. Hence $\symProbability(\symRandomVariate_{1\symHashIndex} < \symRandomVariate_{2\symHashIndex})
=
\symExpRate_1 / (\symExpRate_1 + \symExpRate_2)$ \cite{Mitzenmacher2005}, which finally gives in combination with \eqref{equ:propbability_equal}
\begin{equation*}
\symProbability(\symHashValue_{\symSetA\symHashIndex}
=
\symHashValue_{\symSetB\symHashIndex})
=
\frac{
\sum_{\symUniverseElement\in\symUniverse} \min(\symDiscreteWeight_\symSetA(\symUniverseElement), \symDiscreteWeight_\symSetB(\symUniverseElement))
}{
\sum_{\symUniverseElement\in\symUniverse} \max(\symDiscreteWeight_\symSetA(\symUniverseElement), \symDiscreteWeight_\symSetB(\symUniverseElement))
}
=
\tilde{\symJaccard}.
\end{equation*}
Therefore, signatures defined by \eqref{equ:def_signature} allow unbiased estimation of $\tilde{\symJaccard}$ using \eqref{equ:jaccard_estimator}. 
They can also be used to estimate $\symJaccard$, if the error introduced by the weight discretization can be neglected as discussed in \cref{sec:discretization}. The number of equal components of two signatures follows a binomial distribution with success probability $\tilde{\symJaccard}$, because individual components of the same signature are, like for original minwise hashing, statistically independent by definition.

\cref{alg:simple} is a straightforward implementation of signature definition \eqref{equ:def_signature}. First, the signature components are initialized with infinity. For each pair $(\symUniverseElement, \symWeightIndex)$ a \ac{PRNG} $\symRNG$ is created with the pair as seed to ensure independence. $\symRNG$ is then used to generate $\symHashSize$ exponentially distributed random values which are used to update the signature components. The parameter of $\symRNG$ describes the distribution from which a random value is drawn using $\symRNG$ as random bit source.
Obviously, \cref{alg:simple} is not very efficient, because its time complexity is $\symBigO(\symWeightIndexMax\symHashSize\symInputSize)$. This is much worse than $\symBigO(\symHashSize\symInputSize)$ of the \ac{ICWS} algorithm, because $\symWeightIndexMax$ is expected to be huge. For example, $\symWeightIndexMax$ is in the billions, if weights are discretized using single-precision floating-point numbers. In the oncoming sections we will describe methods to make the calculation of the new signature much more efficient.

The signature components defined by \eqref{equ:def_signature} are all nonnegative real numbers. However, as mentioned in \cref{sec:advanced_techniques}, integer values with a predefined number of bits are often more preferable. Similar to the proposal in \cite{Ioffe2010}, \cref{alg:bbit_extraction} extracts $\symBBit$-bit integer values from the real-valued signature. A \ac{PRNG} is initialized for each signature component with $\symHashValue_{\symHashIndex}$ as seed. It is used to generate a uniform random $\symBBit$-bit integer number which defines the value of the corresponding component of the transformed signature.

\myAlg{
\caption{Straightforward implementation of signature definition \eqref{equ:def_signature}.}
\label{alg:simple}
\KwIn{$\symWeight$}
\KwOut{$\symHashValue_1,
\symHashValue_2,
\ldots,
\symHashValue_{\symHashSize}$}
$(
\symHashValue_1,
\symHashValue_2,
\ldots,
\symHashValue_{\symHashSize}
)\gets
(\infty,\infty,\ldots,\infty)$\;
\ForAll{$\symUniverseElement\in\symUniverse$ such that $\symWeight(\symUniverseElement) \geq \symDiscreteValue_1$}
{
\ForAll{$\symWeightIndex \in \lbrace 1, 2, \ldots, \symWeightIndexMax\rbrace$ such that $\symDiscreteValue_\symWeightIndex\leq\symWeight(\symUniverseElement)$}
{
 $\symRNG \gets$ \New \acs{PRNG} with seed $(\symUniverseElement, \symWeightIndex)$\;
\For{$\symHashIndex \gets 1$ \KwTo $\symHashSize$}{
$\symPoint\gets \symRNG[\text{Exponential}(
\symPropConstant(\symDiscreteValue_\symWeightIndex - \symDiscreteValue_{\symWeightIndex-1}))]$\;
\lIf{$\symPoint < \symHashValue_\symHashIndex$}{$\symHashValue_\symHashIndex \gets \symPoint$}
}
}
}
}

\myAlg{
\caption{Transformation of a real-valued signature into a $\symBBit$-bit hash signature.}
\label{alg:bbit_extraction}
\KwIn{$
\symHashValue_1,
\symHashValue_2,
\ldots,
\symHashValue_{\symHashSize}
$}
\KwOut{$\symHashValue^\symBBit_1, \symHashValue^\symBBit_2, \ldots, \symHashValue^\symBBit_\symHashSize\in\lbrace 0, 1, \ldots, 2^\symBBit-1\rbrace$}
\For{$\symHashIndex \gets 1$ \KwTo $\symHashSize$}{
$\symRNG \gets$ \New \ac{PRNG} with seed $\symHashValue_\symHashIndex$\;
$\symHashValue^\symBBit_\symHashIndex
\gets \symRNG[\text{Uniform}(\lbrace
0,1,\ldots,2^\symBBit-1\rbrace)]$\;
}
}

\subsection{Interpretation as Poisson Process}
\label{sec:interpretation}

Since the random variables $\symPoint_{\symUniverseElement\symWeightIndex\symHashIndex}$ in \eqref{equ:def_signature} are exponentially distributed with rate parameter $\symPropConstant(\symDiscreteValue_\symWeightIndex - \symDiscreteValue_{\symWeightIndex-1})$, they can also be interpreted as first points of independent Poisson processes $\symPoisson_{\symUniverseElement\symWeightIndex\symHashIndex}$ with rates $\symPropConstant(\symDiscreteValue_\symWeightIndex - \symDiscreteValue_{\symWeightIndex-1})$, respectively. Poisson processes have some nice properties. Points can be generated in ascending order, because distances between successive points are independent and exponentially distributed. Moreover, Poisson processes can be combined and split \cite{Mitzenmacher2005}. 

These properties can be exploited to generate the first points of all $\symPoisson_{\symUniverseElement\symWeightIndex\symHashIndex}$
in a different but statistically equivalent way. We consider the combined process $\symPoisson_{\symUniverseElement\symWeightIndex}:=\bigcup_{\symHashIndex=1}^\symHashSize \symPoisson_{\symUniverseElement\symWeightIndex\symHashIndex}$ which has rate $\symPropConstant\symHashSize(\symDiscreteValue_\symWeightIndex - \symDiscreteValue_{\symWeightIndex-1})$. While generating points for $\symPoisson_{\symUniverseElement\symWeightIndex}$ in ascending order, $\symPoisson_{\symUniverseElement\symWeightIndex}$ is split again into the $\symHashSize$ constituent subprocesses $\symPoisson_{\symUniverseElement\symWeightIndex\symHashIndex}$. This can be done by distributing each point of $\symPoisson_{\symUniverseElement\symWeightIndex}$ randomly and uniformly to one of the $\symHashSize$ subprocesses $\symPoisson_{\symUniverseElement\symWeightIndex\symHashIndex}$. As soon as we have assigned at least one point to each of those subprocesses, which means we have found all their first points, we are done. 
Since the constant $\symPropConstant$ is a free parameter, we set $\symPropConstant:=\symHashSize^{-1}$ for simplicity such that $\symPoisson_{\symUniverseElement\symWeightIndex}$ has rate $\symDiscreteValue_\symWeightIndex - \symDiscreteValue_{\symWeightIndex-1}$ in the following.

\cref{alg:enhanced} demonstrates this idea. For each pair $(\symUniverseElement, \symWeightIndex)$ a Poisson process $\symPoisson_{\symUniverseElement\symWeightIndex}$ with rate $\symDiscreteValue_\symWeightIndex - \symDiscreteValue_{\symWeightIndex-1}$ is simulated. A helper class given by \cref{alg:poisson} is used to represent the corresponding state. A new Poisson  process is initialized with starting point $\symPoint = 0$ and with indices $\symWeightIndexMinus = \symWeightIndex-1$ and $\symWeightIndexPlus = \symWeightIndex$ which define the rate as $\symDiscreteValue_\symWeightIndexPlus - \symDiscreteValue_\symWeightIndexMinus$. Furthermore, since the Poisson process object requires an independent random source for point generation, a new \ac{PRNG} is created. By choosing $\symUniverseElement$ and $\symWeightIndex$ as seed we ensure that all processes $\symPoisson_{\symUniverseElement\symWeightIndex}$ are independent. The helper class has a member function \NextPoint{} for generating the next point of the Poisson process by incrementing $\symPoint$ by a value drawn from an exponential distribution with rate parameter $\symDiscreteValue_\symWeightIndexPlus - \symDiscreteValue_\symWeightIndexMinus$. In addition, a uniform random number $\symHashIndex$ drawn from $\lbrace 1, 2, \ldots, \symHashSize\rbrace$ defines the index of the subprocess $\symPoisson_{\symUniverseElement\symWeightIndex\symHashIndex}$ the new point belongs to. 
The helper class also has some more functions as well as an additional field $\symWeight$ for the weight $\symWeight(\symUniverseElement)$ associated with $\symUniverseElement$, which are not yet used in the context of \cref{alg:enhanced}.

\myAlg{
\caption{Improved version of \cref{alg:simple} by generating random values in ascending order for fixed $\symUniverseElement$ and $\symWeightIndex$.}
\label{alg:enhanced}
\KwIn{$\symWeight$}
\KwOut{$\symHashValue_1,
\symHashValue_2,
\ldots,
\symHashValue_{\symHashSize}$}
$(
\symHashValue_1,
\symHashValue_2,
\ldots,
\symHashValue_{\symHashSize}
)\gets
(\infty,\infty,\ldots,\infty)$\;
\ForAll{$\symUniverseElement\in\symUniverse$ such that $\symWeight(\symUniverseElement) \geq \symDiscreteValue_1$}{
\ForAll{$\symWeightIndex \in \lbrace 1, 2, \ldots, \symWeightIndexMax\rbrace$ such that $\symDiscreteValue_{\symWeightIndex}\leq\symWeight(\symUniverseElement)$}{
$\symRNG\gets$ \New\acs{PRNG} with seed $(\symUniverseElement,\symWeightIndex)$\;
$\symPoisson \gets$ \New \PoissonProcess{0, $\symRNG$, $\symWeightIndex-1$, $\symWeightIndex$, $\symWeight(\symUniverseElement)$}\;
$\symPoisson$.\NextPoint{}\;
\While{$\symPoisson.\symPoint \leq \symHashValueMax$}{
\lIf{$\symPoisson.\symPoint < \symHashValue_{\symPoisson.\symHashIndex}$}
{
$\symHashValue_{\symPoisson.\symHashIndex} \gets \symPoisson.\symPoint$}
$\symPoisson$.\NextPoint{}\;
}
}
}
}

\myAlg{
\caption{Helper class for the representation of the combined Poisson process $\symPoisson_{\symUniverseElement,\symWeightIndexMinus:\symWeightIndexPlus} = \bigcup_{\symWeightIndex=\symWeightIndexMinus+1}^\symWeightIndexPlus \symPoisson_{\symUniverseElement\symWeightIndex}
=\bigcup_{\symWeightIndex=\symWeightIndexMinus+1}^\symWeightIndexPlus \bigcup_{\symHashIndex=1}^\symHashSize\symPoisson_{\symUniverseElement\symWeightIndex\symHashIndex}
$.
}
\label{alg:poisson}
\Class{\PoissonProcess}{
\Fields{
$\symPoint$\Com*{current point}
$\symRNG$\Com*{\acl{PRNG}}
$\symWeightIndexMinus$, $\symWeightIndexPlus$\Com*{define the rate as $\symDiscreteValue_\symWeightIndexPlus-\symDiscreteValue_\symWeightIndexMinus$}
$\symWeight$\Com*{weight of $\symUniverseElement$}
$\symHashIndex$\Com*{index of $\symPoisson_{\symUniverseElement\symWeightIndex\symHashIndex}$ point $\symPoint$ is belonging to}
}

\Function(\Com*[f]{constructor, initializes a new object}){
\PoissonProcess{$\symPoint'$, $\symRNG'$, $\symWeightIndexMinusPrime$, $\symWeightIndexPlusPrime$, $\symWeight'$}}{
$\symPoint\gets \symPoint'$\;
$\symRNG\gets\symRNG'$\;
$\symWeightIndexMinus
\gets
\symWeightIndexMinusPrime,\
\symWeightIndexPlus
\gets
\symWeightIndexPlusPrime$\;
$\symWeight\gets\symWeight'$\;
}
\Function(\Com*[f]{generates the next point}){\NextPoint{}}{
$\symPoint\gets
\symPoint + \symRNG[\text{Exponential}(\symDiscreteValue_\symWeightIndexPlus - \symDiscreteValue_\symWeightIndexMinus)]$\;
$\symHashIndex\gets\symRNG[\text{Uniform}(\lbrace 1,2,\ldots, \symHashSize\rbrace)]$\;
}
\Function(\Com*[f]{splits process into two parts}){\Split{}}{
$\symWeightIndexCenter\gets\lfloor(\symWeightIndexMinus+\symWeightIndexPlus)/2\rfloor$\;
$\symRNG'\gets$ \New \acs{PRNG} with seed $(\symPoint, \symWeightIndexCenter)$\;
\eIf{$\symRNG[\text{Bernoulli}(
(\symDiscreteValue_\symWeightIndexCenter - \symDiscreteValue_\symWeightIndexMinus)
/
(\symDiscreteValue_\symWeightIndexPlus - \symDiscreteValue_\symWeightIndexMinus)
)]=1$}{
$\symPoisson'\gets$ \New \PoissonProcess{$\symPoint$, $\symRNG'$, $\symWeightIndexCenter$, $\symWeightIndexPlus$, $\symWeight$}\;
$\symWeightIndexPlus\gets\symWeightIndexCenter$\;
}
{
$\symPoisson'\gets$ \New \PoissonProcess{$\symPoint$, $\symRNG'$, $\symWeightIndexMinus$, $\symWeightIndexCenter$, $\symWeight$}\;
$\symWeightIndexMinus\gets\symWeightIndexCenter$\;
}
\KwRet{$\symPoisson'$}\;
}
\Function{\Splittable{}}{
\KwRet $\symWeightIndexMinus+1 < \symWeightIndexPlus$\;
}
\Function{\PartiallyRelevant{}}{
\KwRet $\symDiscreteValue_{\symWeightIndexMinus+1}\leq\symWeight$\;
}
\Function{\FullyRelevant{}}{
\KwRet $\symDiscreteValue_\symWeightIndexPlus\leq\symWeight$\;
}
}
}

Instead of executing the while-loop until at least one point was generated for all $\symHashSize$ subprocesses $\symPoisson_{\symUniverseElement\symWeightIndex\symHashIndex}$, \cref{alg:enhanced} keeps track of $\symHashValueMax:=\max(
\symHashValue_1,
\symHashValue_2,
\ldots,
\symHashValue_{\symHashSize})$. By definition, points greater than $\symHashValueMax$ are not able to change any of the values $\symHashValue_1,
\symHashValue_2,
\ldots,
\symHashValue_{\symHashSize}$. Therefore, as soon as some point is greater than $\symHashValueMax$ the Poisson process can be stopped. If $\symHashValueMax$ is simultaneously updated each time some signature value $\symHashValue_\symHashIndex$ is replaced by a lower value, $\symHashValueMax$ will decrease over time and the termination condition is satisfied earlier in subsequent iterations of $\symUniverseElement$ and $\symWeightIndex$. 

The efficient maintenance of $\symHashValueMax$ is demonstrated by \cref{alg:track_max}. A binary tree is constructed over $\symHashSize$ leaf nodes with values $\symHashValue_1, \symHashValue_2,\ldots,\symHashValue_\symHashSize$, all starting from positive infinity. The value of a parent node is defined to be the maximum of both children. Therefore, the root node always represents $\symHashValueMax$. If a Poisson process generates some point $\symPoint < \symHashValueMax$, an update of $\symHashValue_\symHashIndex$ and also of its ancestors might be necessary. \cref{alg:track_max} makes a corresponding bottom-up traversal until no further change is necessary.

Conditioned on $\symPoint<\symHashValueMax$ the probability that some node is replaced by a smaller value is less than the reciprocal number of leaves in the corresponding subtree. This means, given $\symPoint<\symHashValueMax$, the probability, that $\symHashValue_\symHashIndex$ is updated, is (of course) at most 1. The probability, that the parent of $\symHashValue_\symHashIndex$ is updated, is at most $\frac{1}{2}$, because it is equally likely that the parent value is given by the sibling of $\symHashValue_\symHashIndex$, and therefore an update of $\symHashValue_\symHashIndex$ will not change the parent. Continuing in this way shows that the expected number of node updates is bounded by the geometric series $1 + 1/2 + 1/4 + \ldots$ and therefore takes amortized constant time and does not contribute to the overall time complexity of \cref{alg:enhanced}. 

Hence, the complexity is primarily dominated by the number of expected \NextPoint{} calls. 
At the beginning, when all $\symHashValue_\symHashIndex$ and therefore also $\symHashValueMax$ are still infinite, the while-loop is executed until all $\symHashValue_\symHashIndex$ have been updated at least once. Since points of $\symPoisson_{\symUniverseElement\symWeightIndex}$ are calculated in ascending order, $\symHashValueMax$ will become finite exactly when the last infinite signature value is overwritten. Thus, $\symHashValueMax$ will also have the same value as the current point of $\symPoisson$ and the termination condition of the while-loop will be immediately satisfied after advancing to the next point using $\symPoisson.$\NextPoint{}. In analogy of the coupon collector's problem \cite{Cormen2009}, the average number of required steps until at least one point is assigned to each signature component will be $\symHashSize\symHarmonic_\symHashSize$ where $\symHarmonic_\symHashSize:=1+\frac{1}{2}+\ldots+\frac{1}{\symHashSize}=\symBigO(\log\symHashSize)$ denotes the $\symHashSize$-th harmonic number.

Dependent on the weight $\symWeight(\symUniverseElement)$ of the input element $\symUniverseElement$, up to $\symWeightIndexMax$ inner for-loop iterations are required. Therefore, the expected number of \NextPoint{} calls needed for the first input element is at most $\symWeightIndexMax(1+\symHashSize\symHarmonic_\symHashSize)$. The number of required cycles will be lower for further elements, because $\symHashValueMax$ decreases with the number of inserted elements and the termination condition will be satisfied earlier. 
The decrease of $\symHashValueMax$ will be faster, if elements with larger weights are processed first, because the corresponding random points are likely to be smaller than for elements with smaller weights. Therefore, it is preferable to process input elements in descending and not in ascending order of weight. 
If we assume random ordering, it can be shown that the expected number of \NextPoint{} calls needed for the $\symInsertionCounter$-th input element is bounded by $\symWeightIndexMax(1+\symHashSize\symHarmonic_\symHashSize/\symInsertionCounter)$. Therefore, the time complexity of \cref{alg:enhanced} is limited by 
$\sum_{\symInsertionCounter=1}^\symInputSize \symWeightIndexMax(1+\symHashSize\symHarmonic_\symHashSize/\symInsertionCounter) = \symBigO(\symWeightIndexMax\symHashSize \log^2 \symHashSize + \symWeightIndexMax \symInputSize)$.
This is already a significant improvement over $\symBigO(\symWeightIndexMax\symHashSize\symInputSize)$ of \cref{alg:simple} for $\symInputSize\gg\symHashSize$ as we have eliminated the factor $\symHashSize$ that accompanies $\symInputSize$. 

However, since $\symWeightIndexMax$ can be very large, further improvements are necessary to compete with $\symBigO(\symHashSize\symInputSize)$ of the \ac{ICWS} algorithm. \cref{alg:enhanced} is only useful for small $\symWeightIndexMax$. In particular, by choosing $\symWeightIndexMax=1$ and $\symDiscreteSet=\lbrace 0, 1\rbrace$ binary weight values can be accurately described. Therefore, 
\cref{alg:enhanced} can be used for the unweighted case and represents an interesting alternative to the advanced minwise hashing algorithms discussed in \cref{sec:advanced_techniques}. The reason is that the time complexity $\symBigO(\symHashSize \log^2 \symHashSize + \symInputSize)$ is comparable to other advanced minwise hashing algorithms discussed in \cref{sec:advanced_techniques}. But, in contrast to them, the signature components are statistically independent like for original minwise hashing.

\myAlg{
\caption{
To efficiently keep track of the global maximum
$\symHashValueMax:=\max(
\symHashValue_1,
\symHashValue_2,
\ldots,
\symHashValue_{\symHashSize})$ while values $\symHashValue_\symHashIndex$ are successively decreased over time, the array is extended by $\symHashValue_{\symHashSize+1},\ldots,\symHashValue_{2\symHashSize-1}$ which correspond to the parent nodes in a binary tree spanned over $\symHashValue_1,
\symHashValue_2,
\ldots,
\symHashValue_{\symHashSize}$. $\symHashValue_{\symHashSize + \lceil \symHashIndex/2 \rceil}$ is the parent of $\symHashValue_\symHashIndex$. A parent node is defined as the maximum of its children $\symHashValue_{\symHashIndex}=\max(\symHashValue_{2(\symHashIndex-\symHashSize)-1}, \symHashValue_{2(\symHashIndex-\symHashSize)})$ for $\symHashIndex > \symHashSize$. If a leaf node $\symHashValue_\symHashIndex$ with $\symHashIndex \leq \symHashSize$ is replaced by a smaller value $\symPoint < \symHashValue_\symHashIndex$, following procedure will be necessary to update the root node $\symHashValue_{2\symHashSize-1}$ which also represents $\symHashValueMax$.
}
\label{alg:track_max}
\KwIn{$\symPoint$, $\symHashIndex$}
$\symHashValue\gets \symPoint$\;
\While{
$\symHashValue < \symHashValue_\symHashIndex$}
{
$\symHashValue_\symHashIndex\gets\symHashValue$\;
$\symHashIndex\gets \symHashSize + \lceil \symHashIndex/2 \rceil$\Com*{calculate parent index}
\lIf{$\symHashIndex\geq 2\symHashSize $}{\Break}
$\symHashValue\gets\max(\symHashValue_{2(\symHashIndex-\symHashSize)-1}, \symHashValue_{2(\symHashIndex-\symHashSize)})$\;
}
}

\subsection{In-Order Point Generation}
The individual treatment of processes $\symPoisson_{\symUniverseElement\symWeightIndex}$ is responsible for the factor $\symWeightIndexMax$ in the time complexity of \cref{alg:enhanced}.
To eliminate or at least to reduce this factor, we need to process all $\symPoisson_{\symUniverseElement\symWeightIndex}$ for a given input element $\symUniverseElement$ in a collective fashion. We consider the combined process $\symPoisson_\symUniverseElement := \bigcup_{\symWeightIndex=1}^\symWeightIndexMax\symPoisson_{\symUniverseElement \symWeightIndex}$ which has rate $\sum_{\symWeightIndex=1}^\symWeightIndexMax \symDiscreteValue_\symWeightIndex-\symDiscreteValue_{\symWeightIndex-1} = \symDiscreteValue_\symWeightIndexMax$. In this context we will call the processes $\symPoisson_{\symUniverseElement \symWeightIndex}$ elementary subprocesses. The idea is now to generate points for $\symPoisson_\symUniverseElement$ in ascending order. To comply with \eqref{equ:def_signature} we must filter points that belong to elementary subprocesses $\symPoisson_{\symUniverseElement \symWeightIndex}$ which satisfy $\symDiscreteValue_\symWeightIndex\leq\symWeight(\symUniverseElement)$. In the following, we will call these elementary subprocesses and also their points relevant.

The properties of Poisson processes allow iterative splitting of $\symPoisson_\symUniverseElement$ until we arrive at some elementary subprocess.
For this, we introduce the notation $\symPoisson_{\symUniverseElement,\symWeightIndexMinus:\symWeightIndexPlus} := \bigcup_{\symWeightIndex=\symWeightIndexMinus+1}^\symWeightIndexPlus \symPoisson_{\symUniverseElement\symWeightIndex}$, which means that $\symPoisson_{\symUniverseElement} = \symPoisson_{\symUniverseElement,0:\symWeightIndexMax}$ and $\symPoisson_{\symUniverseElement \symWeightIndex} = \symPoisson_{\symUniverseElement,\symWeightIndex-1:\symWeightIndex}$.
Assume we have a process $\symPoisson_{\symUniverseElement,\symWeightIndexMinus:\symWeightIndexPlus}$ for which the first point $\symPoint$ has already been generated. If it is not an elementary process, which is equivalent to $\symWeightIndexMinus+1<\symWeightIndexPlus$, it is called splittable and can be divided into two subprocesses $\symPoisson_{\symUniverseElement,\symWeightIndexMinus:\symWeightIndexCenter}$ and $\symPoisson_{\symUniverseElement,\symWeightIndexCenter:\symWeightIndexPlus}$ with
$\symWeightIndexMinus<\symWeightIndexCenter<\symWeightIndexPlus$. The probability that $\symPoint$ is a point of $\symPoisson_{\symUniverseElement,\symWeightIndexMinus:\symWeightIndexCenter}$ or $\symPoisson_{\symUniverseElement,\symWeightIndexCenter:\symWeightIndexPlus}$ is proportional to the corresponding rates, respectively. Hence, the result of a Bernoulli trial can be used to decide to which subprocess the point $\symPoint$ actually belongs to. The first point of the other subprocess can be generated by adding to $\symPoint$ a random number drawn from an exponential distribution with rate parameter equal to the rate of that subprocess.

\cref{alg:poisson} shows how the splitting operation can be realized. The \Split{} method first calculates the split index as $\symWeightIndexCenter=\lfloor(\symWeightIndexMinus+\symWeightIndexPlus)/2\rfloor$, such that the weight index range is divided into two parts of almost equal size. A Bernoulli trial with success probability $(\symDiscreteValue_\symWeightIndexCenter - \symDiscreteValue_\symWeightIndexMinus)
/
(\symDiscreteValue_\symWeightIndexPlus - \symDiscreteValue_\symWeightIndexMinus)$ determines whether $\symPoint$ is a point of 
$\symPoisson_{\symUniverseElement,\symWeightIndexMinus:\symWeightIndexCenter}$.
If not, $\symPoint$ must belong to $\symPoisson_{\symUniverseElement,\symWeightIndexCenter:\symWeightIndexPlus}$. The parent process object can be reused for the elected subprocess by simply adjusting either $\symWeightIndexPlus$ or $\symWeightIndexMinus$, respectively. A new process object $\symPoisson'$ is needed to represent the second subprocess. It is initialized with starting point $\symPoint$ and a new \ac{PRNG}. Independence can be ensured by using $\symPoint$ together with the splitting index $\symWeightIndexCenter$ as seed. The splitting index makes the seed unique, because further splittings are not possible with same index. The object of the second child $\symPoisson'$ is finally returned by the \Split{} method. A call to $\symPoisson'\!.$\NextPoint{} must follow later to get the first point of $\symPoisson'$. 

In order to calculate the relevant points of $\symPoisson_\symUniverseElement$ in ascending order, we first generate the very first point of 
$\symPoisson_\symUniverseElement$. Next, we determine, to which elementary subprocess the point belongs to. The corresponding subprocess can be found through iterative splitting, always proceeding with the child that contains the point of the parent process. This procedure corresponds to searching for the smallest point in a binary tree, where each node is equal to the minimum of both children. The next smallest point of $\symPoisson_\symUniverseElement$ can be found analogously. Among all subprocesses, which resulted from all splittings so far, 
we choose the subprocess with smallest current point and restart the same splitting procedure as before from there. In this way all points of $\symPoisson_\symUniverseElement$ and its associated elementary subprocesses $\symPoisson_{\symUniverseElement\symWeightIndex}$ can be found in ascending order.

Since we are only interested in relevant points, we must omit irrelevant ones. However, $\symDiscreteValue_\symWeightIndexMax$ is often much larger than $\symWeight(\symUniverseElement)$, which makes point-wise filtering not very efficient. Luckily, the hierarchical structure allows efficient skipping of irrelevant points. 
A subprocess $\symPoisson_{\symUniverseElement,\symWeightIndexMinus:\symWeightIndexPlus}$ only contains relevant points, if at least one of its elementary processes is relevant, which is the case if $\symDiscreteValue_{\symWeightIndexMinus+1}\leq\symWeight(\symUniverseElement)$. In this case, we say $\symPoisson_{\symUniverseElement,\symWeightIndexMinus:\symWeightIndexPlus}$ is partially relevant. If $\symPoisson_{\symUniverseElement,\symWeightIndexMinus:\symWeightIndexPlus}$ is not partially relevant, it does not contain any relevant points and can immediately be discarded without further processing. 
We also define a subprocess $\symPoisson_{\symUniverseElement,\symWeightIndexMinus:\symWeightIndexPlus}$ to be fully relevant, if all its elementary subprocesses and corresponding points are relevant, which is the case if $\symDiscreteValue_{\symWeightIndexPlus}\leq\symWeight(\symUniverseElement)$. Full relevance implies partial relevance. If a process is an elementary subprocess and therefore not further splittable, partial and full relevance will be equivalent.

\myAlg{
\caption{BagMinHash~1.}
\label{alg:bagminhash1}
\KwIn{$\symWeight$}
\KwOut{$\symHashValue_1,
\symHashValue_2,
\ldots,
\symHashValue_{\symHashSize}$}
$(
\symHashValue_1,
\symHashValue_2,
\ldots,
\symHashValue_{\symHashSize}
)\gets
(\infty,\infty,\ldots,\infty)$\;
\ForAll{$\symUniverseElement\in\symUniverse$ such that $\symWeight(\symUniverseElement) \geq \symDiscreteValue_1$}{
$\symRNG\gets$ \New\acs{PRNG} with seed $\symUniverseElement$\;
$\symPoisson \gets$ \New \PoissonProcess{0, $\symRNG$, 0, $\symWeightIndexMax$, $\symWeight(\symUniverseElement)$}\;
$\symPoisson.$\NextPoint{}\;
\lIf{$\symPoisson.$\FullyRelevant{} \And $\symPoisson.\symPoint < \symHashValue_{\symPoisson.\symHashIndex}$}{
$\symHashValue_{\symPoisson.\symHashIndex} \gets \symPoisson.\symPoint$}
create new min-heap for Poisson processes $\symPoisson$ with $\symPoisson.\symPoint$ as key\;
\While{$\symPoisson.\symPoint \leq \symHashValueMax$}
{
\While{
$\symPoisson.$\Splittable{} \And $
\symPoisson.$\PartiallyRelevant{}}
{
$\symPoisson' \gets \symPoisson.$\Split{}\;
\lIf{$\symPoisson.$\FullyRelevant{} \And $\symPoisson.\symPoint < \symHashValue_{\symPoisson.\symHashIndex}$}{
$\symHashValue_{\symPoisson.\symHashIndex} \gets \symPoisson.\symPoint$}
\If{$\symPoisson'\!.$\PartiallyRelevant{}}
{
$\symPoisson'\!.$\NextPoint{}\;
\lIf{$\symPoisson'\!.$\FullyRelevant{} \And $\symPoisson'\!.\symPoint < \symHashValue_{\symPoisson'\!.\symHashIndex}$}{
$\symHashValue_{\symPoisson'\!.\symHashIndex} \gets \symPoisson'\!.\symPoint$}
\lIf{$\symPoisson'\!.\symPoint \leq \symHashValueMax$}{
push $\symPoisson'$ to heap
}
}
}
\If{$\symPoisson.$\FullyRelevant{}}{
$\symPoisson$.\NextPoint{}\;
\lIf{$\symPoisson.\symPoint < \symHashValue_{\symPoisson.\symHashIndex}$}{
$\symHashValue_{\symPoisson.\symHashIndex} \gets \symPoisson.\symPoint$}
\lIf{$\symPoisson.\symPoint \leq \symHashValueMax$}{
push $\symPoisson$ to heap
}
}
\lIf{heap is empty}{\Break}
$\symPoisson\gets$ pop from heap\;
}
}
}

\cref{alg:bagminhash1} called BagMinHash~1 implements the described procedure. The algorithm loops over all elements having nonzero discrete weight ($\symWeight(\symUniverseElement)\geq \symDiscreteValue_1$). 
For each element $\symUniverseElement$ a new Poisson process object $\symPoisson$ is created that represents $\symPoisson_\symUniverseElement$. The weight $\symWeight(\symUniverseElement)$ is passed to the Poisson process object, which is needed by the \PartiallyRelevant{} and \FullyRelevant{} member functions. 
After generating the first point and updating the corresponding signature value in case $\symPoisson$ is already fully relevant, a while-loop is started. Therein, the process is split as long as possible and as long as it is partially relevant. After each split operation, $\symPoisson$ and also its sibling $\symPoisson'$ are checked for full relevance. If this is the case, the corresponding signature value $\symHashValue_{\symPoisson.\symHashIndex}$ or $\symHashValue_{\symPoisson'\!.\symHashIndex}$ and if necessary also $\symHashValueMax$ are updated. These early updates allow leaving the outer while-loop as soon as possible. 
If the sibling $\symPoisson'$ is partially relevant, it needs to be processed later and must be saved in the meantime. However, since the current point of $\symPoisson'$ represents a lower bound for all its further points, $\symPoisson'$ can be discarded as a whole, if $\symPoisson'\!.\symPoint$ is already larger than $\symHashValueMax$. Otherwise, we put $\symPoisson'$ into a min-heap with $\symPoisson'\!.\symPoint$ as key. In this way the subprocess with smallest current point can be efficiently retrieved later.

The inner while-loop can be left due to two reasons: First, if $\symPoisson$ is no longer partially relevant, which may be the consequence of the last split operation, we can immediately discard $\symPoisson$. Second, if $\symPoisson$ is no longer splittable and corresponds to an elementary subprocess, while $\symPoisson$ is still partially relevant, it must also be fully relevant. In this case, we generate the next point of $\symPoisson$, update the corresponding signature value if necessary, and push it into the min-heap. In both cases we continue with the process having the smallest current point, which can be retrieved from the min-heap. Once the point of the current process is larger than $\symHashValueMax$, which also means that all remaining points of any other processes are larger, we are finished as further signature value updates can be ruled out.

The time complexity of this algorithm can be  bounded by 
$\symBigO(\symComplexityFunc(\symHashSize, \symWeightIndexMax) + (\log\symWeightIndexMax)(\log\log\symWeightIndexMax) \symInputSize)$ where $\symComplexityFunc$ denotes a function independent of $\symInputSize$. The reason is that for large data sizes $\symInputSize\gg\symHashSize$, point generation can be stopped most of the time after finding the smallest relevant point. The corresponding binary search over all $\symWeightIndexMax$ elementary subprocesses explains the $\log \symWeightIndexMax$ factor. The min-heap insertions and removals contribute the $\log\log \symWeightIndexMax$ factor. Hence, if $\log\symWeightIndexMax$ is much smaller than $\symHashSize$, the new approach is potentially much faster than \ac{ICWS}. 

\subsection{Optimization}
\label{sec:optimization}

\myAlg{
\caption{BagMinHash~2, an optimized variant of \cref{alg:bagminhash1}.}
\label{alg:bagminhash2}
\KwIn{$\symWeight$}
\KwOut{$\symHashValue_1,
\symHashValue_2,
\ldots,
\symHashValue_{\symHashSize}$}
$(
\symHashValue_1,
\symHashValue_2,
\ldots,
\symHashValue_{\symHashSize}
)\gets
(\infty,\infty,\ldots,\infty)$\;
create empty buffer\;
\ForAll{$\symUniverseElement\in\symUniverse$ such that $\symWeight(\symUniverseElement) \geq \symDiscreteValue_1$}{
create new min-heap for Poisson processes $\symPoisson$ with $\symPoisson.\symPoint$ as key\;
$\symRNG\gets$ \New\acs{PRNG} with seed $\symUniverseElement$\;
$\symPoisson \gets$ \New \PoissonProcess{0, $\symRNG$, 0, $\symWeightIndexMax$, $\symWeight(\symUniverseElement)$}\;
$\symPoisson.$\NextPoint{}\;
\lIf{$\symPoisson.$\FullyRelevant{} \And $\symPoisson.\symPoint < \symHashValue_{\symPoisson.\symHashIndex}$}{
$\symHashValue_{\symPoisson.\symHashIndex} \gets \symPoisson.\symPoint$}
\While{$\symPoisson.\symPoint \leq \symHashValueMax$}
{
\While{
$\symPoisson.$\Splittable{} \And $
\symPoisson.$\PartiallyRelevant{} \And \Not $
\symPoisson.$\FullyRelevant{}}
{
$\symPoisson' \gets \symPoisson.$\Split{}\;
\lIf{$\symPoisson.$\FullyRelevant{} \And $\symPoisson.\symPoint < \symHashValue_{\symPoisson.\symHashIndex}$}{
$\symHashValue_{\symPoisson.\symHashIndex} \gets \symPoisson.\symPoint$}
\If{$\symPoisson'\!.$\PartiallyRelevant{}}
{
$\symPoisson'\!.$\NextPoint{}\;
\lIf{$\symPoisson'\!.$\FullyRelevant{} \And $\symPoisson'\!.\symPoint < \symHashValue_{\symPoisson'\!.\symHashIndex}$}{
$\symHashValue_{\symPoisson'\!.\symHashIndex} \gets \symPoisson'\!.\symPoint$}
\lIf{$\symPoisson'\!.\symPoint \leq \symHashValueMax$}{
push $\symPoisson'$ to heap
}
}
}
\If{$\symPoisson.$\FullyRelevant{}}{
push $\symPoisson$ to heap\;
\Break\;
}
\lIf{heap is empty}{\Break}
$\symPoisson\gets$ pop from heap\;
}

add all heap elements $\symPoisson$ with $\symPoisson.\symPoint \leq \symHashValueMax$ to buffer\;
}
create new min-heap from all buffer elements $\symPoisson$ with $\symPoisson.\symPoint \leq \symHashValueMax$\;
\While{\Not heap is empty}
{
$\symPoisson\gets$ pop from heap\;
\lIf{$\symPoisson.\symPoint > \symHashValueMax$}{\Break}
\While{
$\symPoisson.$\Splittable{} \And $
\symPoisson.$\PartiallyRelevant{}}
{
$\symPoisson' \gets \symPoisson.$\Split{}\;
\lIf{$\symPoisson.$\FullyRelevant{} \And $\symPoisson.\symPoint < \symHashValue_{\symPoisson.\symHashIndex}$}{
$\symHashValue_{\symPoisson.\symHashIndex} \gets \symPoisson.\symPoint$}
\If{$\symPoisson'\!.$\PartiallyRelevant{}}
{
$\symPoisson'\!.$\NextPoint{}\;
\lIf{$\symPoisson'\!.$\FullyRelevant{} \And $\symPoisson'\!.\symPoint < \symHashValue_{\symPoisson'\!.\symHashIndex}$}{
$\symHashValue_{\symPoisson'\!.\symHashIndex} \gets \symPoisson'\!.\symPoint$}
\lIf{$\symPoisson'\!.\symPoint \leq \symHashValueMax$}{
push $\symPoisson'$ to heap
}
}
}
\If{$\symPoisson.$\FullyRelevant{}}{
$\symPoisson$.\NextPoint{}\;
\lIf{$\symPoisson.\symPoint < \symHashValue_{\symPoisson.\symHashIndex}$}{
$\symHashValue_{\symPoisson.\symHashIndex} \gets \symPoisson.\symPoint$}
\lIf{$\symPoisson.\symPoint \leq \symHashValueMax$}{
push $\symPoisson$ to heap
}
}
}
}

The new algorithm can be further optimized. The processing time for inserting a new element $\symUniverseElement$ decreases with the number of already inserted elements, because $\symHashValueMax$ is continuously decreasing. In particular, the processing time of the very first element takes a lot of time, because at least one point needs to be found for each signature position. However, if there are much more input elements to come, it is very likely that most of these signature values are overwritten by points from subsequent input elements. Therefore, we can save some computation time, if we divide the processing of individual input elements into two phases. In a first phase, we calculate only the smallest relevant point of $\symPoisson_\symUniverseElement$ for each input element $\symUniverseElement$ and store any subprocesses, which result from splittings and which have first points not greater than $\symHashValueMax$, in a buffer for later processing. If there are enough input elements $\symInputSize \gg \symHashSize$, $\symHashValueMax$ will already be very small after the first step. In a second phase, we put all buffered processes into a new min-heap, which allows processing them in order of their current points. Subprocesses of further splittings are again pushed into this min-heap. 
As soon as $\symHashValueMax$ is smaller than the first element of the min-heap, the signature computation is finished.

BagMinHash~2 shown as \cref{alg:bagminhash2} implements the described optimization and is logically equivalent to BagMinHash~1. The first for-loop represents the first phase of the improved approach. As soon as the smallest relevant point is found for an input element $\symUniverseElement$, which is the case once the first fully relevant subprocess has been found, the iterative splitting process is stopped. All subprocesses, that have been forked from $\symPoisson_\symUniverseElement$ so far, are stored in a buffer before continuing with the next input element.
The second step starts with the construction of a min-heap filled with all the subprocesses that have been collected in the buffer while processing all input elements in the first phase. Then, these subprocesses are processed as in \cref{alg:bagminhash1} until all remaining processes have points larger than $\symHashValueMax$.

In order to keep space requirements for the buffer small, it can be realized as max-heap. In this way, since $\symHashValueMax$ is decreasing  during the first step, subprocesses with points greater than $\symHashValueMax$ can be immediately evicted in an efficient way.

\section{Experimental Results}
For our tests we have implemented both variants of the BagMinHash algorithm using C++\footnote{Source code that has been used to produce the results presented in this paper is made available on \url{https://github.com/oertl/bagminhash}.}. Since BagMinHash assumes an ideal \ac{PRNG}, its implementation requires some special care. We have chosen to use the xxHash64 hashing algorithm \cite{Collet2016}, which we applied to the given seed in combination with values from a predefined integer sequence to get as many different 64-bit random numbers as needed. To keep the number of hash function evaluations small, we tried to use as few random bits as possible to generate random numbers for different distributions as required by BagMinHash. Bernoulli trials are realized using an algorithm that takes only two random  bits on average \cite{Devroye1986}. Uniform discrete values can be efficiently generated as described in \cite{Lumbroso2013}. Moreover, we used the ziggurat algorithm to produce exponentially distributed random numbers without the need of costly logarithm evaluations \cite{Marsaglia2000}.

In our experiments we primarily used the set of nonnegative single-precision floating point numbers as described in \cref{sec:discretization} for weight discretization. However, double-precision is used for all calculations and the representation of points of Poisson processes.
Moreover, we assumed that all input elements are 64-bit integers. This corresponds to a universe $\symUniverse$ with dimensionality $|\symUniverse| = 2^{64}$, which is large enough for most real-world applications. Any bag from a different universe can be mapped to a bag of 64-bit integers by hashing its elements.

\subsection{Verification}

\begin{table*}
\caption{Analysis of the estimation error of various weighted minwise hashing algorithms for different test cases and different hash sizes.
The empirical \aclp{MSE} $\symEmpiricalMSE$ and corresponding $\symZScore$-scores have been calculated from \num{10000} random examples, respectively. $\symEmpiricalMSE$ values with magnitude greater than 3 indicate exceptional behavior and are colored red.}
\label{tbl:error}
\centering
\resizebox{\textwidth}{!}{
\begin{tabular}{lrrrrrrrrrrrrrrrr}
\toprule
& &
& \multicolumn{2}{c}{BagMinHash (float)}
& \multicolumn{2}{c}{BagMinHash (binary)}
& \multicolumn{2}{c}{\acs*{ICWS} \cite{Ioffe2010}}
& \multicolumn{2}{c}{0-bit \cite{Li2015}}
& \multicolumn{2}{c}{\acs*{CCWS} \cite{Wu2016}}
& \multicolumn{2}{c}{\acs*{PCWS} \cite{Wu2017a}}
& \multicolumn{2}{c}{\acs*{I2CWS} \cite{Wu2017}}
\\
\cmidrule(l){4-5}
\cmidrule(l){6-7}
\cmidrule(l){8-9}
\cmidrule(l){10-11}
\cmidrule(l){12-13}
\cmidrule(l){14-15}
\cmidrule(l){16-17}
test case & \symHashSize & $\symExpectation(\symEmpiricalMSE)$
& $\symEmpiricalMSE$ & $\symZScore$-score
& $\symEmpiricalMSE$ & $\symZScore$-score
& $\symEmpiricalMSE$ & $\symZScore$-score
& $\symEmpiricalMSE$ & $\symZScore$-score
& $\symEmpiricalMSE$ & $\symZScore$-score
& $\symEmpiricalMSE$ & $\symZScore$-score
& $\symEmpiricalMSE$ & $\symZScore$-score
\\
\midrule
\multirowcell{4}[1em][l]{$\lbrace(1,10)\rbrace$ \\ $\symJaccard = \num[group-digits = false]{0.1}$}
& 4
& \numsci{ 2.25E-02}
&
\numsci{ 2.19E-02}
&
\num{ -1.49}
&
N/A
&
N/A
&
\numsci{ 2.23E-02}
&
\num{ -0.43}
&
\numsci{ 8.10E-01}
&
\color{red}\bf
\numsci{ 1.93E+03}
&
\numsci{ 1.00E-02}
&
\color{red}\bf
\numsci{ -3.07E+01}
&
\numsci{ 2.25E-02}
&
\num{ -0.04}
&
\numsci{ 2.23E-02}
&
\num{ -0.44}
\\
& 16
& \numsci{ 5.63E-03}
&
\numsci{ 5.75E-03}
&
\num{ 1.44}
&
N/A
&
N/A
&
\numsci{ 5.55E-03}
&
\num{ -0.91}
&
\numsci{ 8.10E-01}
&
\color{red}\bf
\numsci{ 9.39E+03}
&
\numsci{ 1.00E-02}
&
\color{red}\bf
\numsci{ 5.11E+01}
&
\numsci{ 5.67E-03}
&
\num{ 0.55}
&
\numsci{ 5.70E-03}
&
\num{ 0.82}
\\
& 64
& \numsci{ 1.41E-03}
&
\numsci{ 1.42E-03}
&
\num{ 0.75}
&
N/A
&
N/A
&
\numsci{ 1.38E-03}
&
\num{ -1.30}
&
\numsci{ 8.10E-01}
&
\color{red}\bf
\numsci{ 3.99E+04}
&
\numsci{ 1.00E-02}
&
\color{red}\bf
\numsci{ 4.24E+02}
&
\numsci{ 1.41E-03}
&
\num{ 0.12}
&
\numsci{ 1.42E-03}
&
\num{ 0.76}
\\
& 256
& \numsci{ 3.52E-04}
&
\numsci{ 3.48E-04}
&
\num{ -0.66}
&
N/A
&
N/A
&
\numsci{ 3.50E-04}
&
\num{ -0.36}
&
\numsci{ 8.10E-01}
&
\color{red}\bf
\numsci{ 1.62E+05}
&
\numsci{ 1.00E-02}
&
\color{red}\bf
\numsci{ 1.93E+03}
&
\numsci{ 3.59E-04}
&
\num{ 1.45}
&
\numsci{ 3.42E-04}
&
\num{ -1.94}
\\
& 1024
& \numsci{ 8.79E-05}
&
\numsci{ 8.85E-05}
&
\num{ 0.50}
&
N/A
&
N/A
&
\numsci{ 8.91E-05}
&
\num{ 1.00}
&
\numsci{ 8.10E-01}
&
\color{red}\bf
\numsci{ 6.51E+05}
&
\numsci{ 1.00E-02}
&
\color{red}\bf
\numsci{ 7.96E+03}
&
\numsci{ 8.95E-05}
&
\num{ 1.28}
&
\numsci{ 8.56E-05}
&
\num{ -1.82}
\\
& 4096
& \numsci{ 2.20E-05}
&
\numsci{ 2.26E-05}
&
\num{ 1.91}
&
N/A
&
N/A
&
\numsci{ 2.18E-05}
&
\num{ -0.44}
&
\numsci{ 8.10E-01}
&
\color{red}\bf
\numsci{ 2.61E+06}
&
\numsci{ 1.00E-02}
&
\color{red}\bf
\numsci{ 3.21E+04}
&
\numsci{ 2.18E-05}
&
\num{ -0.65}
&
\numsci{ 2.19E-05}
&
\num{ -0.09}
\\
\midrule
\multirowcell{4}[1em][l]{$\lbrace(9,10)\rbrace$ \\ $\symJaccard = \num[group-digits = false]{0.9}$}
& 4
& \numsci{ 2.25E-02}
&
\numsci{ 2.27E-02}
&
\num{ 0.40}
&
N/A
&
N/A
&
\numsci{ 2.24E-02}
&
\num{ -0.33}
&
\numsci{ 1.00E-02}
&
\color{red}\bf
\numsci{ -3.07E+01}
&
\numsci{ 8.10E-01}
&
\color{red}\bf
\numsci{ 1.93E+03}
&
\numsci{ 2.25E-02}
&
\num{ 0.02}
&
\numsci{ 2.28E-02}
&
\num{ 0.77}
\\
& 16
& \numsci{ 5.62E-03}
&
\numsci{ 5.77E-03}
&
\num{ 1.66}
&
N/A
&
N/A
&
\numsci{ 5.66E-03}
&
\num{ 0.46}
&
\numsci{ 1.00E-02}
&
\color{red}\bf
\numsci{ 5.11E+01}
&
\numsci{ 8.10E-01}
&
\color{red}\bf
\numsci{ 9.39E+03}
&
\numsci{ 5.66E-03}
&
\num{ 0.42}
&
\numsci{ 5.60E-03}
&
\num{ -0.35}
\\
& 64
& \numsci{ 1.41E-03}
&
\numsci{ 1.39E-03}
&
\num{ -1.00}
&
N/A
&
N/A
&
\numsci{ 1.36E-03}
&
\num{ -2.15}
&
\numsci{ 1.00E-02}
&
\color{red}\bf
\numsci{ 4.24E+02}
&
\numsci{ 8.10E-01}
&
\color{red}\bf
\numsci{ 3.99E+04}
&
\numsci{ 1.38E-03}
&
\num{ -1.21}
&
\numsci{ 1.41E-03}
&
\num{ 0.19}
\\
& 256
& \numsci{ 3.52E-04}
&
\numsci{ 3.55E-04}
&
\num{ 0.63}
&
N/A
&
N/A
&
\numsci{ 3.50E-04}
&
\num{ -0.40}
&
\numsci{ 1.00E-02}
&
\color{red}\bf
\numsci{ 1.93E+03}
&
\numsci{ 8.10E-01}
&
\color{red}\bf
\numsci{ 1.62E+05}
&
\numsci{ 3.54E-04}
&
\num{ 0.55}
&
\numsci{ 3.52E-04}
&
\num{ 0.07}
\\
& 1024
& \numsci{ 8.79E-05}
&
\numsci{ 8.63E-05}
&
\num{ -1.30}
&
N/A
&
N/A
&
\numsci{ 8.70E-05}
&
\num{ -0.76}
&
\numsci{ 1.00E-02}
&
\color{red}\bf
\numsci{ 7.96E+03}
&
\numsci{ 8.10E-01}
&
\color{red}\bf
\numsci{ 6.51E+05}
&
\numsci{ 9.02E-05}
&
\num{ 1.87}
&
\numsci{ 8.79E-05}
&
\num{ -0.02}
\\
& 4096
& \numsci{ 2.20E-05}
&
\numsci{ 2.20E-05}
&
\num{ 0.11}
&
N/A
&
N/A
&
\numsci{ 2.16E-05}
&
\num{ -1.13}
&
\numsci{ 1.00E-02}
&
\color{red}\bf
\numsci{ 3.21E+04}
&
\numsci{ 8.10E-01}
&
\color{red}\bf
\numsci{ 2.61E+06}
&
\numsci{ 2.20E-05}
&
\num{ 0.24}
&
\numsci{ 2.18E-05}
&
\num{ -0.66}
\\
\midrule
\multirowcell{4}[1em][l]{$\lbrace(3,20),(30,7)\rbrace$ \\ $\symJaccard = \num[group-digits = false]{0.2}$}
& 4
& \numsci{ 4.00E-02}
&
\numsci{ 4.01E-02}
&
\num{ 0.09}
&
N/A
&
N/A
&
\numsci{ 4.07E-02}
&
\num{ 1.19}
&
\numsci{ 8.04E-02}
&
\color{red}\bf
\numsci{ 7.04E+01}
&
\numsci{ 4.00E-02}
&
\num{ 0.00}
&
\numsci{ 3.45E-02}
&
\color{red}\bf
\num{ -9.50}
&
\numsci{ 3.32E-02}
&
\color{red}\bf
\numsci{ -1.18E+01}
\\
& 16
& \numsci{ 1.00E-02}
&
\numsci{ 1.00E-02}
&
\num{ -0.00}
&
N/A
&
N/A
&
\numsci{ 9.88E-03}
&
\num{ -0.83}
&
\numsci{ 3.73E-02}
&
\color{red}\bf
\numsci{ 1.92E+02}
&
\numsci{ 4.00E-02}
&
\color{red}\bf
\numsci{ 2.11E+02}
&
\numsci{ 1.09E-02}
&
\color{red}\bf
\num{ 6.48}
&
\numsci{ 2.00E-02}
&
\color{red}\bf
\numsci{ 7.05E+01}
\\
& 64
& \numsci{ 2.50E-03}
&
\numsci{ 2.51E-03}
&
\num{ 0.35}
&
N/A
&
N/A
&
\numsci{ 2.52E-03}
&
\num{ 0.53}
&
\numsci{ 2.63E-02}
&
\color{red}\bf
\numsci{ 6.72E+02}
&
\numsci{ 4.00E-02}
&
\color{red}\bf
\numsci{ 1.06E+03}
&
\numsci{ 4.87E-03}
&
\color{red}\bf
\numsci{ 6.69E+01}
&
\numsci{ 1.69E-02}
&
\color{red}\bf
\numsci{ 4.07E+02}
\\
& 256
& \numsci{ 6.25E-04}
&
\numsci{ 6.21E-04}
&
\num{ -0.51}
&
N/A
&
N/A
&
\numsci{ 6.38E-04}
&
\num{ 1.47}
&
\numsci{ 2.35E-02}
&
\color{red}\bf
\numsci{ 2.59E+03}
&
\numsci{ 4.00E-02}
&
\color{red}\bf
\numsci{ 4.45E+03}
&
\numsci{ 3.47E-03}
&
\color{red}\bf
\numsci{ 3.22E+02}
&
\numsci{ 1.62E-02}
&
\color{red}\bf
\numsci{ 1.76E+03}
\\
& 1024
& \numsci{ 1.56E-04}
&
\numsci{ 1.55E-04}
&
\num{ -0.73}
&
N/A
&
N/A
&
\numsci{ 1.58E-04}
&
\num{ 0.86}
&
\numsci{ 2.28E-02}
&
\color{red}\bf
\numsci{ 1.02E+04}
&
\numsci{ 4.00E-02}
&
\color{red}\bf
\numsci{ 1.80E+04}
&
\numsci{ 3.08E-03}
&
\color{red}\bf
\numsci{ 1.32E+03}
&
\numsci{ 1.59E-02}
&
\color{red}\bf
\numsci{ 7.13E+03}
\\
& 4096
& \numsci{ 3.91E-05}
&
\numsci{ 3.87E-05}
&
\num{ -0.60}
&
N/A
&
N/A
&
\numsci{ 3.89E-05}
&
\num{ -0.35}
&
\numsci{ 2.26E-02}
&
\color{red}\bf
\numsci{ 4.08E+04}
&
\numsci{ 4.00E-02}
&
\color{red}\bf
\numsci{ 7.23E+04}
&
\numsci{ 2.99E-03}
&
\color{red}\bf
\numsci{ 5.34E+03}
&
\numsci{ 1.59E-02}
&
\color{red}\bf
\numsci{ 2.87E+04}
\\
\midrule
\multirowcell{4}[1em][l]{$\lbrace(0,2),(3,4),(6,3),(2,4)\rbrace$ \\ $\symJaccard = \num[group-digits = false]{0.5}$}
& 4
& \numsci{ 6.25E-02}
&
\numsci{ 6.20E-02}
&
\num{ -0.66}
&
N/A
&
N/A
&
\numsci{ 6.41E-02}
&
\num{ 2.07}
&
\numsci{ 6.82E-02}
&
\color{red}\bf
\num{ 7.47}
&
\numsci{ 2.50E-01}
&
\color{red}\bf
\numsci{ 2.45E+02}
&
\numsci{ 6.74E-02}
&
\color{red}\bf
\num{ 6.34}
&
\numsci{ 8.17E-02}
&
\color{red}\bf
\numsci{ 2.51E+01}
\\
& 16
& \numsci{ 1.56E-02}
&
\numsci{ 1.54E-02}
&
\num{ -0.83}
&
N/A
&
N/A
&
\numsci{ 1.54E-02}
&
\num{ -1.08}
&
\numsci{ 2.19E-02}
&
\color{red}\bf
\numsci{ 2.96E+01}
&
\numsci{ 2.50E-01}
&
\color{red}\bf
\numsci{ 1.10E+03}
&
\numsci{ 2.09E-02}
&
\color{red}\bf
\numsci{ 2.47E+01}
&
\numsci{ 3.86E-02}
&
\color{red}\bf
\numsci{ 1.07E+02}
\\
& 64
& \numsci{ 3.91E-03}
&
\numsci{ 3.80E-03}
&
\num{ -1.90}
&
N/A
&
N/A
&
\numsci{ 3.88E-03}
&
\num{ -0.43}
&
\numsci{ 1.10E-02}
&
\color{red}\bf
\numsci{ 1.29E+02}
&
\numsci{ 2.50E-01}
&
\color{red}\bf
\numsci{ 4.49E+03}
&
\numsci{ 8.76E-03}
&
\color{red}\bf
\numsci{ 8.86E+01}
&
\numsci{ 2.78E-02}
&
\color{red}\bf
\numsci{ 4.37E+02}
\\
& 256
& \numsci{ 9.77E-04}
&
\numsci{ 9.84E-04}
&
\num{ 0.53}
&
N/A
&
N/A
&
\numsci{ 9.64E-04}
&
\num{ -0.89}
&
\numsci{ 8.12E-03}
&
\color{red}\bf
\numsci{ 5.19E+02}
&
\numsci{ 2.50E-01}
&
\color{red}\bf
\numsci{ 1.81E+04}
&
\numsci{ 5.91E-03}
&
\color{red}\bf
\numsci{ 3.58E+02}
&
\numsci{ 2.53E-02}
&
\color{red}\bf
\numsci{ 1.76E+03}
\\
& 1024
& \numsci{ 2.44E-04}
&
\numsci{ 2.44E-04}
&
\num{ -0.15}
&
N/A
&
N/A
&
\numsci{ 2.42E-04}
&
\num{ -0.49}
&
\numsci{ 7.51E-03}
&
\color{red}\bf
\numsci{ 2.11E+03}
&
\numsci{ 2.50E-01}
&
\color{red}\bf
\numsci{ 7.24E+04}
&
\numsci{ 5.25E-03}
&
\color{red}\bf
\numsci{ 1.45E+03}
&
\numsci{ 2.46E-02}
&
\color{red}\bf
\numsci{ 7.05E+03}
\\
& 4096
& \numsci{ 6.10E-05}
&
\numsci{ 6.18E-05}
&
\num{ 0.91}
&
N/A
&
N/A
&
\numsci{ 6.06E-05}
&
\num{ -0.55}
&
\numsci{ 7.33E-03}
&
\color{red}\bf
\numsci{ 8.42E+03}
&
\numsci{ 2.50E-01}
&
\color{red}\bf
\numsci{ 2.90E+05}
&
\numsci{ 5.09E-03}
&
\color{red}\bf
\numsci{ 5.83E+03}
&
\numsci{ 2.44E-02}
&
\color{red}\bf
\numsci{ 2.82E+04}
\\
\midrule
\multirowcell{4}[1em][l]{$\lbrace(4,2)^{15},(1,4)^{10},(12,0)^{5}\rbrace$ \\ $\symJaccard = \num[group-digits = false]{0.25}$}
& 4
& \numsci{ 4.69E-02}
&
\numsci{ 4.73E-02}
&
\num{ 0.60}
&
N/A
&
N/A
&
\numsci{ 4.66E-02}
&
\num{ -0.45}
&
\numsci{ 4.69E-02}
&
\num{ 0.08}
&
\numsci{ 6.25E-02}
&
\color{red}\bf
\numsci{ 2.46E+01}
&
\numsci{ 4.30E-02}
&
\color{red}\bf
\num{ -6.10}
&
\numsci{ 4.40E-02}
&
\color{red}\bf
\num{ -4.52}
\\
& 16
& \numsci{ 1.17E-02}
&
\numsci{ 1.17E-02}
&
\num{ -0.22}
&
N/A
&
N/A
&
\numsci{ 1.17E-02}
&
\num{ -0.18}
&
\numsci{ 1.17E-02}
&
\num{ 0.16}
&
\numsci{ 6.25E-02}
&
\color{red}\bf
\numsci{ 3.10E+02}
&
\numsci{ 1.18E-02}
&
\num{ 0.77}
&
\numsci{ 2.54E-02}
&
\color{red}\bf
\numsci{ 8.36E+01}
\\
& 64
& \numsci{ 2.93E-03}
&
\numsci{ 2.99E-03}
&
\num{ 1.46}
&
N/A
&
N/A
&
\numsci{ 2.93E-03}
&
\num{ 0.03}
&
\numsci{ 3.03E-03}
&
\num{ 2.44}
&
\numsci{ 6.25E-02}
&
\color{red}\bf
\numsci{ 1.44E+03}
&
\numsci{ 4.55E-03}
&
\color{red}\bf
\numsci{ 3.92E+01}
&
\numsci{ 2.04E-02}
&
\color{red}\bf
\numsci{ 4.24E+02}
\\
& 256
& \numsci{ 7.32E-04}
&
\numsci{ 7.24E-04}
&
\num{ -0.82}
&
N/A
&
N/A
&
\numsci{ 7.32E-04}
&
\num{ 0.01}
&
\numsci{ 8.08E-04}
&
\color{red}\bf
\num{ 7.33}
&
\numsci{ 6.25E-02}
&
\color{red}\bf
\numsci{ 5.97E+03}
&
\numsci{ 2.63E-03}
&
\color{red}\bf
\numsci{ 1.84E+02}
&
\numsci{ 1.92E-02}
&
\color{red}\bf
\numsci{ 1.78E+03}
\\
& 1024
& \numsci{ 1.83E-04}
&
\numsci{ 1.84E-04}
&
\num{ 0.43}
&
N/A
&
N/A
&
\numsci{ 1.82E-04}
&
\num{ -0.62}
&
\numsci{ 2.53E-04}
&
\color{red}\bf
\numsci{ 2.69E+01}
&
\numsci{ 6.25E-02}
&
\color{red}\bf
\numsci{ 2.41E+04}
&
\numsci{ 2.13E-03}
&
\color{red}\bf
\numsci{ 7.52E+02}
&
\numsci{ 1.89E-02}
&
\color{red}\bf
\numsci{ 7.21E+03}
\\
& 4096
& \numsci{ 4.58E-05}
&
\numsci{ 4.53E-05}
&
\num{ -0.68}
&
N/A
&
N/A
&
\numsci{ 4.52E-05}
&
\num{ -0.90}
&
\numsci{ 1.15E-04}
&
\color{red}\bf
\numsci{ 1.08E+02}
&
\numsci{ 6.25E-02}
&
\color{red}\bf
\numsci{ 9.65E+04}
&
\numsci{ 1.99E-03}
&
\color{red}\bf
\numsci{ 3.01E+03}
&
\numsci{ 1.88E-02}
&
\color{red}\bf
\numsci{ 2.89E+04}
\\
\midrule
\multirowcell{4}[1em][l]{$\bigcup_{\symTestCaseIndex=0}^{1000} \lbrace({1.001}^\symTestCaseIndex, {1.002}^{\symTestCaseIndex})\rbrace$ \\ $\symJaccard = \num[group-digits = false]{0.538308}$}
& 4
& \numsci{ 6.21E-02}
&
\numsci{ 6.14E-02}
&
\num{ -1.02}
&
N/A
&
N/A
&
\numsci{ 6.18E-02}
&
\num{ -0.37}
&
\numsci{ 6.14E-02}
&
\num{ -0.94}
&
\numsci{ 2.84E-01}
&
\color{red}\bf
\numsci{ 2.90E+02}
&
\numsci{ 6.75E-02}
&
\color{red}\bf
\num{ 6.99}
&
\numsci{ 1.02E-01}
&
\color{red}\bf
\numsci{ 5.28E+01}
\\
& 16
& \numsci{ 1.55E-02}
&
\numsci{ 1.57E-02}
&
\num{ 0.98}
&
N/A
&
N/A
&
\numsci{ 1.56E-02}
&
\num{ 0.53}
&
\numsci{ 1.55E-02}
&
\num{ -0.33}
&
\numsci{ 2.82E-01}
&
\color{red}\bf
\numsci{ 1.25E+03}
&
\numsci{ 1.98E-02}
&
\color{red}\bf
\numsci{ 2.03E+01}
&
\numsci{ 5.99E-02}
&
\color{red}\bf
\numsci{ 2.09E+02}
\\
& 64
& \numsci{ 3.88E-03}
&
\numsci{ 3.78E-03}
&
\num{ -1.90}
&
N/A
&
N/A
&
\numsci{ 3.88E-03}
&
\num{ -0.02}
&
\numsci{ 3.83E-03}
&
\num{ -0.97}
&
\numsci{ 2.81E-01}
&
\color{red}\bf
\numsci{ 5.09E+03}
&
\numsci{ 7.93E-03}
&
\color{red}\bf
\numsci{ 7.42E+01}
&
\numsci{ 4.97E-02}
&
\color{red}\bf
\numsci{ 8.41E+02}
\\
& 256
& \numsci{ 9.71E-04}
&
\numsci{ 9.82E-04}
&
\num{ 0.79}
&
N/A
&
N/A
&
\numsci{ 9.68E-04}
&
\num{ -0.22}
&
\numsci{ 9.60E-04}
&
\num{ -0.82}
&
\numsci{ 2.81E-01}
&
\color{red}\bf
\numsci{ 2.04E+04}
&
\numsci{ 4.96E-03}
&
\color{red}\bf
\numsci{ 2.91E+02}
&
\numsci{ 4.72E-02}
&
\color{red}\bf
\numsci{ 3.37E+03}
\\
& 1024
& \numsci{ 2.43E-04}
&
\numsci{ 2.48E-04}
&
\num{ 1.40}
&
N/A
&
N/A
&
\numsci{ 2.36E-04}
&
\num{ -1.81}
&
\numsci{ 2.36E-04}
&
\num{ -2.07}
&
\numsci{ 2.81E-01}
&
\color{red}\bf
\numsci{ 8.19E+04}
&
\numsci{ 4.31E-03}
&
\color{red}\bf
\numsci{ 1.19E+03}
&
\numsci{ 4.67E-02}
&
\color{red}\bf
\numsci{ 1.35E+04}
\\
& 4096
& \numsci{ 6.07E-05}
&
\numsci{ 6.09E-05}
&
\num{ 0.30}
&
N/A
&
N/A
&
\numsci{ 6.10E-05}
&
\num{ 0.43}
&
\numsci{ 6.10E-05}
&
\num{ 0.36}
&
\numsci{ 2.81E-01}
&
\color{red}\bf
\numsci{ 3.28E+05}
&
\numsci{ 4.12E-03}
&
\color{red}\bf
\numsci{ 4.74E+03}
&
\numsci{ 4.66E-02}
&
\color{red}\bf
\numsci{ 5.42E+04}
\\
\midrule
\multirowcell{4}[1em][l]{$\lbrace(0,1),(1,0),(1,1)\rbrace$ \\ $\symJaccard = \num[group-digits = false]{0.333333}$}
& 4
& \numsci{ 5.56E-02}
&
\numsci{ 5.53E-02}
&
\num{ -0.41}
&
\numsci{ 5.59E-02}
&
\num{ 0.46}
&
\numsci{ 5.50E-02}
&
\num{ -0.74}
&
\numsci{ 5.57E-02}
&
\num{ 0.24}
&
\numsci{ 5.63E-02}
&
\num{ 1.02}
&
\numsci{ 5.67E-02}
&
\num{ 1.57}
&
\numsci{ 5.60E-02}
&
\num{ 0.60}
\\
& 16
& \numsci{ 1.39E-02}
&
\numsci{ 1.39E-02}
&
\num{ 0.06}
&
\numsci{ 1.39E-02}
&
\num{ -0.11}
&
\numsci{ 1.41E-02}
&
\num{ 0.95}
&
\numsci{ 1.37E-02}
&
\num{ -1.05}
&
\numsci{ 1.41E-02}
&
\num{ 1.29}
&
\numsci{ 1.39E-02}
&
\num{ -0.20}
&
\numsci{ 1.38E-02}
&
\num{ -0.44}
\\
& 64
& \numsci{ 3.47E-03}
&
\numsci{ 3.44E-03}
&
\num{ -0.58}
&
\numsci{ 3.48E-03}
&
\num{ 0.06}
&
\numsci{ 3.50E-03}
&
\num{ 0.66}
&
\numsci{ 3.42E-03}
&
\num{ -1.05}
&
\numsci{ 3.43E-03}
&
\num{ -0.81}
&
\numsci{ 3.45E-03}
&
\num{ -0.52}
&
\numsci{ 3.61E-03}
&
\num{ 2.85}
\\
& 256
& \numsci{ 8.68E-04}
&
\numsci{ 8.70E-04}
&
\num{ 0.17}
&
\numsci{ 8.55E-04}
&
\num{ -1.09}
&
\numsci{ 8.84E-04}
&
\num{ 1.33}
&
\numsci{ 8.60E-04}
&
\num{ -0.64}
&
\numsci{ 8.76E-04}
&
\num{ 0.65}
&
\numsci{ 8.77E-04}
&
\num{ 0.73}
&
\numsci{ 8.66E-04}
&
\num{ -0.13}
\\
& 1024
& \numsci{ 2.17E-04}
&
\numsci{ 2.13E-04}
&
\num{ -1.24}
&
\numsci{ 2.13E-04}
&
\num{ -1.45}
&
\numsci{ 2.25E-04}
&
\num{ 2.45}
&
\numsci{ 2.20E-04}
&
\num{ 0.86}
&
\numsci{ 2.18E-04}
&
\num{ 0.20}
&
\numsci{ 2.21E-04}
&
\num{ 1.24}
&
\numsci{ 2.15E-04}
&
\num{ -0.54}
\\
& 4096
& \numsci{ 5.43E-05}
&
\numsci{ 5.50E-05}
&
\num{ 0.98}
&
\numsci{ 5.36E-05}
&
\num{ -0.80}
&
\numsci{ 5.44E-05}
&
\num{ 0.21}
&
\numsci{ 5.43E-05}
&
\num{ 0.08}
&
\numsci{ 5.52E-05}
&
\num{ 1.27}
&
\numsci{ 5.39E-05}
&
\num{ -0.44}
&
\numsci{ 5.43E-05}
&
\num{ 0.02}
\\
\midrule
\multirowcell{4}[1em][l]{$\lbrace(0,1)^{30},(1,0)^{10},(1,1)^{160}\rbrace$ \\ $\symJaccard = \num[group-digits = false]{0.8}$}
& 4
& \numsci{ 4.00E-02}
&
\numsci{ 3.98E-02}
&
\num{ -0.39}
&
\numsci{ 3.92E-02}
&
\num{ -1.47}
&
\numsci{ 4.00E-02}
&
\num{ -0.07}
&
\numsci{ 3.97E-02}
&
\num{ -0.44}
&
\numsci{ 4.02E-02}
&
\num{ 0.37}
&
\numsci{ 4.02E-02}
&
\num{ 0.37}
&
\numsci{ 4.02E-02}
&
\num{ 0.29}
\\
& 16
& \numsci{ 1.00E-02}
&
\numsci{ 9.95E-03}
&
\num{ -0.33}
&
\numsci{ 1.00E-02}
&
\num{ 0.33}
&
\numsci{ 1.00E-02}
&
\num{ 0.28}
&
\numsci{ 9.81E-03}
&
\num{ -1.33}
&
\numsci{ 9.81E-03}
&
\num{ -1.34}
&
\numsci{ 9.95E-03}
&
\num{ -0.34}
&
\numsci{ 1.01E-02}
&
\num{ 0.93}
\\
& 64
& \numsci{ 2.50E-03}
&
\numsci{ 2.55E-03}
&
\num{ 1.28}
&
\numsci{ 2.42E-03}
&
\num{ -2.32}
&
\numsci{ 2.56E-03}
&
\num{ 1.81}
&
\numsci{ 2.45E-03}
&
\num{ -1.48}
&
\numsci{ 2.50E-03}
&
\num{ 0.12}
&
\numsci{ 2.48E-03}
&
\num{ -0.63}
&
\numsci{ 2.52E-03}
&
\num{ 0.58}
\\
& 256
& \numsci{ 6.25E-04}
&
\numsci{ 6.19E-04}
&
\num{ -0.66}
&
\numsci{ 6.16E-04}
&
\num{ -1.02}
&
\numsci{ 6.23E-04}
&
\num{ -0.25}
&
\numsci{ 6.17E-04}
&
\num{ -0.89}
&
\numsci{ 6.31E-04}
&
\num{ 0.64}
&
\numsci{ 6.07E-04}
&
\num{ -2.00}
&
\numsci{ 6.31E-04}
&
\num{ 0.68}
\\
& 1024
& \numsci{ 1.56E-04}
&
\numsci{ 1.54E-04}
&
\num{ -1.10}
&
\numsci{ 1.55E-04}
&
\num{ -0.40}
&
\numsci{ 1.57E-04}
&
\num{ 0.18}
&
\numsci{ 1.56E-04}
&
\num{ 0.05}
&
\numsci{ 1.55E-04}
&
\num{ -0.54}
&
\numsci{ 1.52E-04}
&
\num{ -2.07}
&
\numsci{ 1.57E-04}
&
\num{ 0.24}
\\
& 4096
& \numsci{ 3.91E-05}
&
\numsci{ 3.94E-05}
&
\num{ 0.66}
&
\numsci{ 3.85E-05}
&
\num{ -0.98}
&
\numsci{ 3.89E-05}
&
\num{ -0.25}
&
\numsci{ 3.95E-05}
&
\num{ 0.86}
&
\numsci{ 3.93E-05}
&
\num{ 0.43}
&
\numsci{ 3.92E-05}
&
\num{ 0.27}
&
\numsci{ 3.82E-05}
&
\num{ -1.53}
\\
\midrule
\multirowcell{4}[1em][l]{$\lbrace(0,1)^{300},(1,0)^{500},(1,1)^{1200}\rbrace$ \\ $\symJaccard = \num[group-digits = false]{0.6}$}
& 4
& \numsci{ 6.00E-02}
&
\numsci{ 5.97E-02}
&
\num{ -0.38}
&
\numsci{ 5.96E-02}
&
\num{ -0.56}
&
\numsci{ 5.92E-02}
&
\num{ -1.05}
&
\numsci{ 5.86E-02}
&
\num{ -1.90}
&
\numsci{ 6.07E-02}
&
\num{ 0.88}
&
\numsci{ 6.13E-02}
&
\num{ 1.78}
&
\numsci{ 6.02E-02}
&
\num{ 0.21}
\\
& 16
& \numsci{ 1.50E-02}
&
\numsci{ 1.52E-02}
&
\num{ 0.97}
&
\numsci{ 1.51E-02}
&
\num{ 0.29}
&
\numsci{ 1.49E-02}
&
\num{ -0.43}
&
\numsci{ 1.49E-02}
&
\num{ -0.45}
&
\numsci{ 1.47E-02}
&
\num{ -1.36}
&
\numsci{ 1.51E-02}
&
\num{ 0.71}
&
\numsci{ 1.50E-02}
&
\num{ -0.05}
\\
& 64
& \numsci{ 3.75E-03}
&
\numsci{ 3.78E-03}
&
\num{ 0.52}
&
\numsci{ 3.83E-03}
&
\num{ 1.49}
&
\numsci{ 3.70E-03}
&
\num{ -0.90}
&
\numsci{ 3.73E-03}
&
\num{ -0.38}
&
\numsci{ 3.75E-03}
&
\num{ -0.01}
&
\numsci{ 3.81E-03}
&
\num{ 1.18}
&
\numsci{ 3.73E-03}
&
\num{ -0.33}
\\
& 256
& \numsci{ 9.37E-04}
&
\numsci{ 9.30E-04}
&
\num{ -0.54}
&
\numsci{ 9.53E-04}
&
\num{ 1.17}
&
\numsci{ 9.52E-04}
&
\num{ 1.08}
&
\numsci{ 9.36E-04}
&
\num{ -0.12}
&
\numsci{ 9.36E-04}
&
\num{ -0.10}
&
\numsci{ 9.36E-04}
&
\num{ -0.14}
&
\numsci{ 9.32E-04}
&
\num{ -0.41}
\\
& 1024
& \numsci{ 2.34E-04}
&
\numsci{ 2.32E-04}
&
\num{ -0.70}
&
\numsci{ 2.37E-04}
&
\num{ 0.79}
&
\numsci{ 2.34E-04}
&
\num{ -0.00}
&
\numsci{ 2.32E-04}
&
\num{ -0.58}
&
\numsci{ 2.40E-04}
&
\num{ 1.55}
&
\numsci{ 2.33E-04}
&
\num{ -0.56}
&
\numsci{ 2.33E-04}
&
\num{ -0.35}
\\
& 4096
& \numsci{ 5.86E-05}
&
\numsci{ 5.76E-05}
&
\num{ -1.24}
&
\numsci{ 5.93E-05}
&
\num{ 0.81}
&
\numsci{ 5.89E-05}
&
\num{ 0.41}
&
\numsci{ 5.78E-05}
&
\num{ -0.99}
&
\numsci{ 5.89E-05}
&
\num{ 0.32}
&
\numsci{ 5.76E-05}
&
\num{ -1.25}
&
\numsci{ 5.95E-05}
&
\num{ 1.12}
\\
\bottomrule
\end{tabular}

}
\end{table*}

For the verification of the BagMinHash algorithm we used synthetic data. The reason is that it is difficult to extract many pairs of input vectors from real data sets that have some predefined Jaccard index. Furthermore, a couple of recently published algorithms with theoretical flaws have been wrongly justified by tests on real-world data. As we will see these algorithms did not pass our tests, which is an indication that our tests are quite selective.

Each of our test cases is characterized by a bag of weight pairs $\bigcup_{\symTestCaseIndex}\lbrace(\symWeight_{\symSetA\symTestCaseIndex},\symWeight_{\symSetB\symTestCaseIndex})\rbrace$. Each pair stands for some unique input element $\symUniverseElement$ which has weight $\symWeight_\symSetA(\symUniverseElement) = \symWeight_{\symSetA\symTestCaseIndex}$ in the first bag $\symSetA$ and weight $\symWeight_\symSetB(\symUniverseElement) = \symWeight_{\symSetB\symTestCaseIndex}$ in the second bag $\symSetB$, respectively. Since we are free to choose the elements that are associated with the weight pairs, we can simply draw 64-bit random numbers. In this way we can generate as many pairs of bags as needed for our evaluations. All of them will have the same Jaccard similarity. To avoid that elements are processed in some particular order, the elements of a bag are always shuffled before hashing.

\cref{tbl:error} lists all test cases together with the corresponding Jaccard similarity, which we used to test our new BagMinHash algorithm and to compare it to other weighted minwise hashing algorithms. We also varied the signature size $\symHashSize\in\lbrace 4, 16, 64, 256, 1024, 4096\rbrace$. For each test case $\symNumExamples=\num{10000}$ pairs of bags have been generated. 
For each pair an estimate $\symJaccardEstimator_\symSomeIndex$ of the Jaccard similarity was obtained using the corresponding hash signatures and \eqref{equ:jaccard_estimator}. We calculated the empirical \ac{MSE} as
$\symEmpiricalMSE = 
\frac{1}{\symNumExamples}
\sum_{\symSomeIndex=1}^\symNumExamples
(\symJaccardEstimator_\symSomeIndex - \symJaccard)^2$. Since hash collisions can be ignored for 64-bit signature values, the number of matching signature positions follows a binomial distribution with success probability $\symJaccard$. The expectation and the variance of $\symEmpiricalMSE$ can be derived as 
$\symExpectation(\symEmpiricalMSE) = \frac{\symJaccard(1-\symJaccard)}{\symHashSize}$ and $\symVariance(\symEmpiricalMSE) = 
\frac{\symJaccard^2(1-\symJaccard)^2}{\symHashSize^2\symNumExamples}
\left(
2-\frac{6}{\symHashSize}
\right)
+
\frac{\symJaccard(1-\symJaccard)}{\symHashSize^3\symNumExamples}$, respectively. 
These formulas can be used to normalize $\symEmpiricalMSE$, which yields the corresponding $\symZScore$-score
\begin{equation*}
\text{$\symZScore$-score} = 
\left(\symEmpiricalMSE - \symExpectation(\symEmpiricalMSE)\right)/\sqrt{\symVariance(\symEmpiricalMSE)}.
\end{equation*}

\cref{tbl:error} shows the empirical \acp{MSE} and the corresponding $\symZScore$-scores of our simulations for various algorithms. 
In case of BagMinHash we got identical results for both variants, which is expected as both variants are logically equivalent. The results do not show any evidence for unexpected behavior, because all $\symZScore$-scores have magnitudes smaller than 3. Values with larger magnitudes would have been shown in red color.
We also applied BagMinHash with binary weight discretization ($\symDiscreteSet = \lbrace 0, 1\rbrace$, $\symWeightIndexMax=1$) to test cases with only weights equal to 0 or 1.
Again, no anomalies have been observed and the theoretically predicted \acp{MSE} have been confirmed. As expected, the \ac{ICWS} also passes all our test cases. 

Much more interesting are a couple of algorithms that have been recently presented as supposed improvements over \ac{ICWS}. They share the same time complexity $\symBigO(\symHashSize\symInputSize)$ of the \ac{ICWS} algorithm, but introduce some modifications to save a couple of operations. As example, 0-bit hashing \cite{Li2015} was proposed to simplify the generation of integer signature values and to avoid a post-processing step like \cref{alg:bbit_extraction}. It was reported that the resulting signatures approximate \eqref{equ:signature} reasonably well in practice. Unfortunately, a rigorous analysis of the validity range of this approximation was not made. Our results show that the algorithm has problems for $\symInputSize < \symHashSize$ except for test cases with binary weights.

There is also a series of recently published modifications of \ac{ICWS} algorithm that did not pass our tests either: 
Canonical consistent weighted sampling \acused{CCWS}(\ac{CCWS}) \cite{Wu2016}, \ac{PCWS} \cite{Wu2017a}, and \ac{I2CWS} \cite{Wu2017}. When looking for the reason, we discovered theoretical flaws described in the following using the notation of the corresponding papers. In \cite{Wu2016} the samples of $y_k$ are drawn from $[S_k - r_k, S_k]$ instead from $[0, S_k]$ which, however, is necessary for consistent sampling. In \cite{Wu2017a} the expression $\operatorname{pdf}(y, a) = \frac{1}{S}(yu_1^{-1})e^{-(yu_1^{-1})a}$ is turned into $\operatorname{pdf}(y, a) = \frac{1}{S}(Se^{-Sa})$ by replacing $yu_1^{-1}$ by its expectation $\operatorname{\mathbb{E}}(yu_1^{-1})=S$ which is an invalid transformation. And finally, in \cite{Wu2017} 
the proof of consistency assumes that $t^S_{k_\ast 2} =  t^T_{k_\ast 2}$ follows from $y_{k_\ast} \leq T_{k_\ast} \leq S_{k_\ast}$, which is not true either. Incorrect argumentation even led to the claim that the \ac{ICWS} algorithm is wrong, which we could neither confirm theoretically nor empirically.

\subsection{Performance}

\begin{figure}[t]
\centering
\subfloat{\includegraphics[width = 0.49\columnwidth]{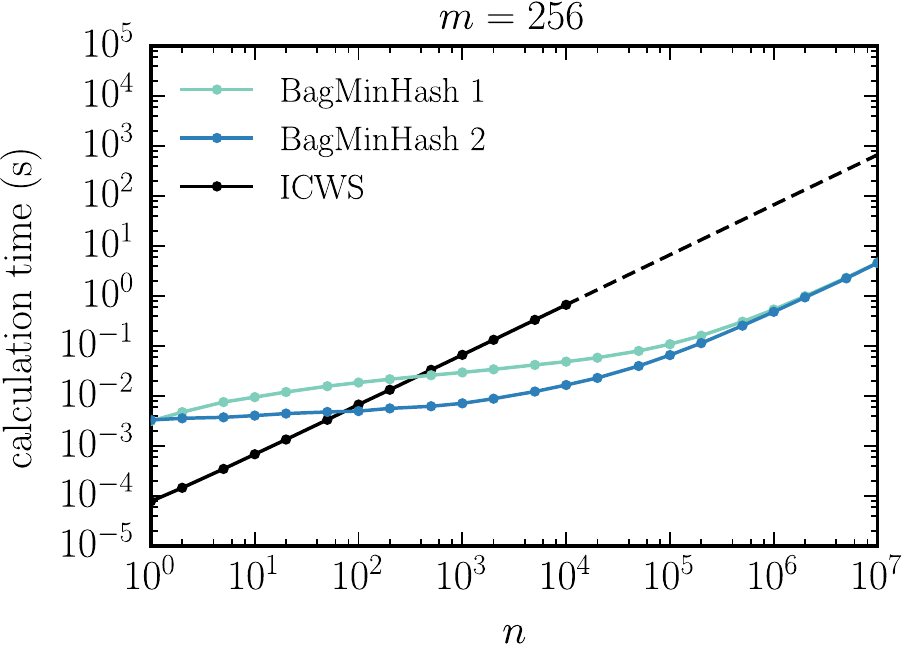}} 
\hfill
\subfloat{\includegraphics[width = 0.49\columnwidth]{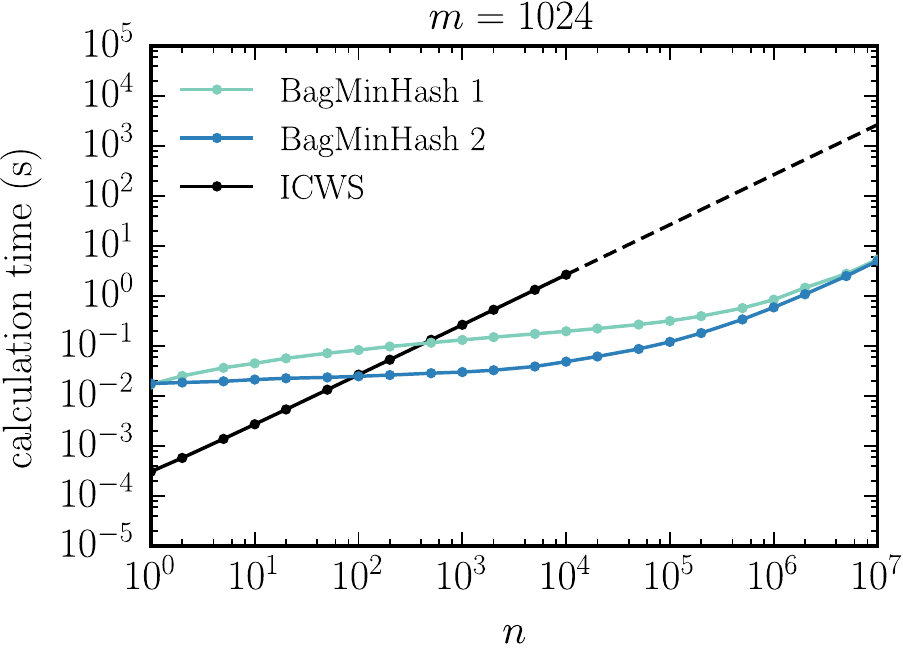}}\\
\subfloat{\includegraphics[width = 0.49\columnwidth]{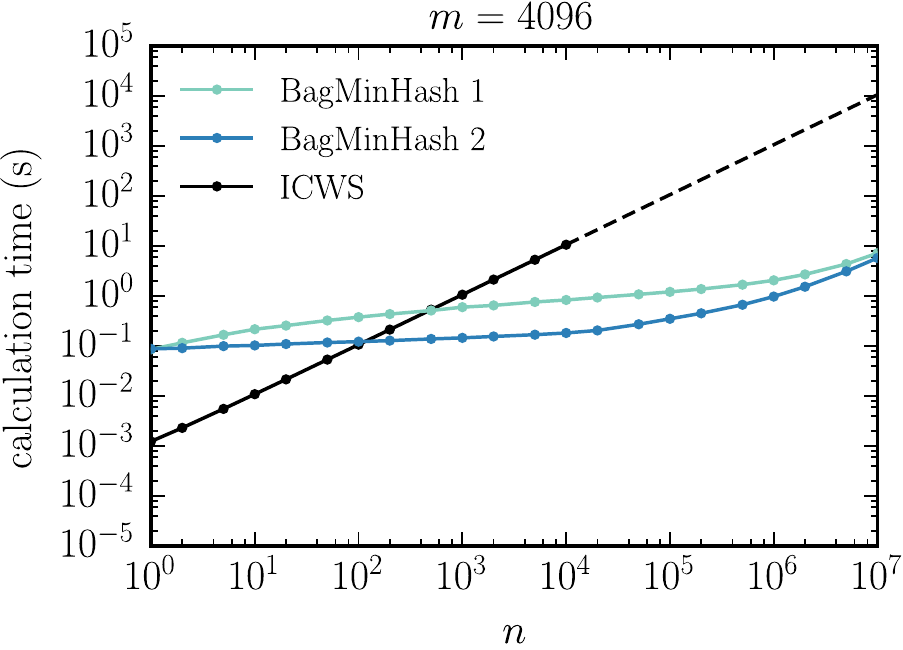}} 
\hfill
\subfloat{\includegraphics[width = 0.49\columnwidth]{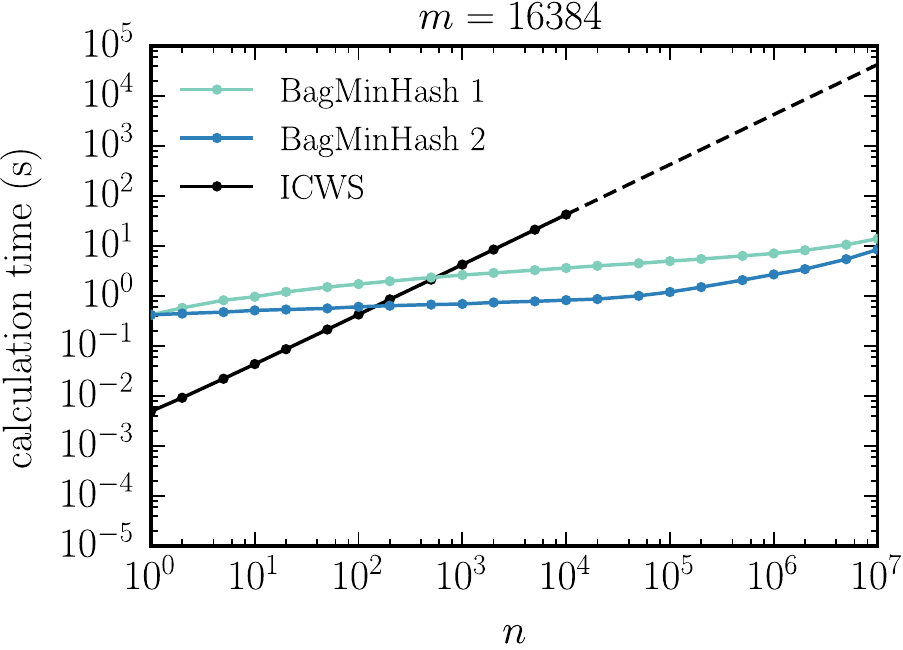}}
\caption{Average calculation time for random bags of size $\symInputSize$ with weights drawn from an exponential distribution with rate 1.}
\label{fig:speed}
\end{figure}

We still need to check if BagMinHash is really fast in practice. Therefore, we measured the performance for various signature sizes $\symHashSize\in\lbrace 256, 1024, 4096, 16384\rbrace$ and bag sizes $\symInputSize$ ranging from $1$ to \num{e7}. For each case we randomly generated \num{100} bags of 64-bit random integer numbers with weights drawn from an exponential distribution with rate parameter 1.
The average calculation times on an Intel Core i5-2500K CPU for both BagMinHash variants and \ac{ICWS} are shown in \cref{fig:speed}. 
The calculation time of \ac{ICWS} is extrapolated for $\symInputSize>10000$ by utilizing its perfect proportional scaling with $\symInputSize$.

\begin{figure}[t]
\centering
\subfloat{\includegraphics[width = 0.49\columnwidth]{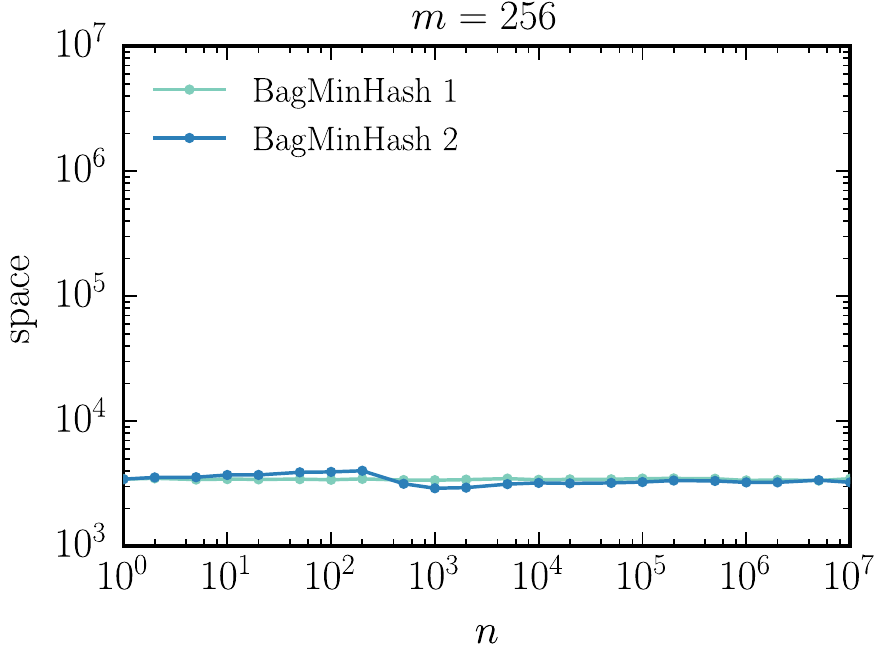}} 
\hfill
\subfloat{\includegraphics[width = 0.49\columnwidth]{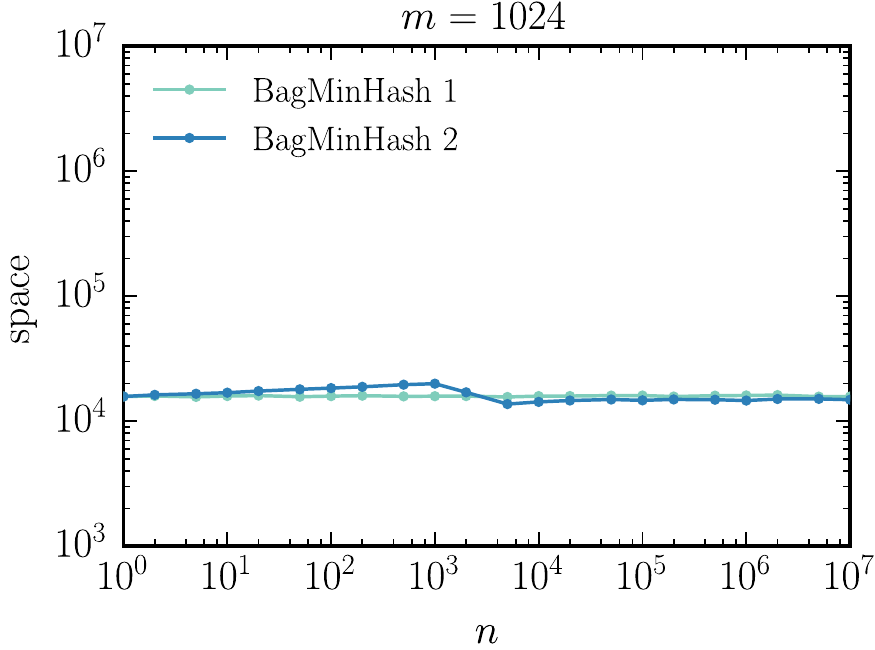}}\\
\subfloat{\includegraphics[width = 0.49\columnwidth]{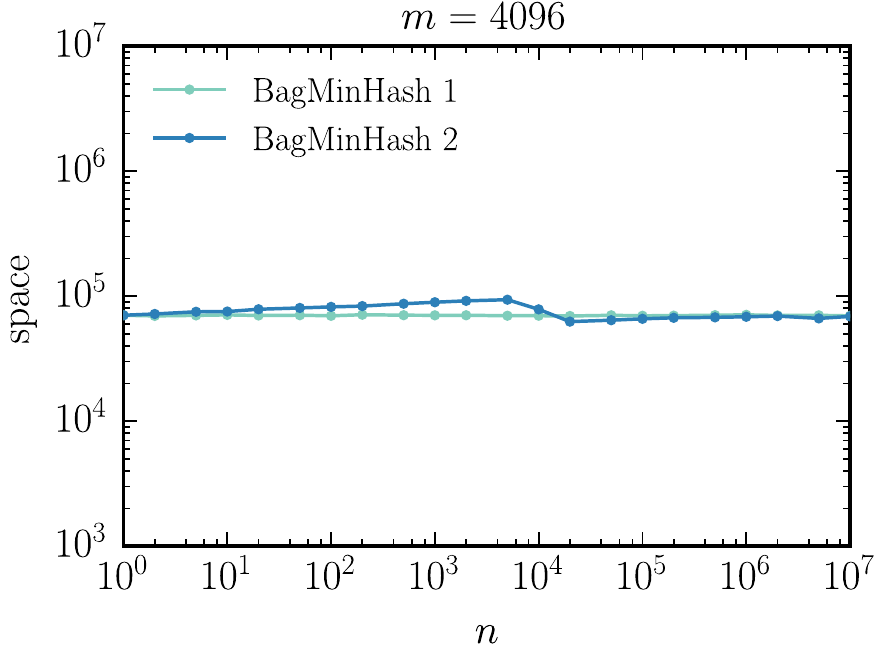}} 
\hfill
\subfloat{\includegraphics[width = 0.49\columnwidth]{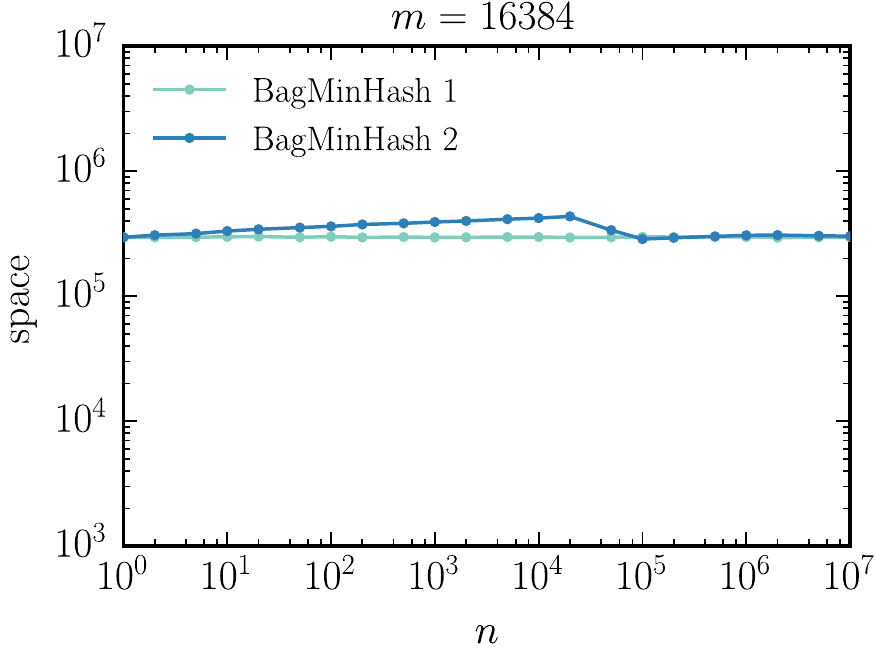}}
\caption{The average space requirements in terms of stored Poisson process objects as defined in \cref{alg:poisson}.}
\label{fig:space}
\end{figure}

The BagMinHash algorithms are significantly slower than \ac{ICWS} for small $\symInputSize$. 
The reason can be seen when considering the case $\symInputSize=1$ for which, on average, the BagMinHash algorithms require the generation of $\symHashSize\symHarmonic_\symHashSize = \symBigO(\symHashSize\log\symHashSize)$ points until at least one point is found for every  signature component as discussed in \cref{sec:interpretation}. Furthermore, the iterative splitting of the Poisson process to generate one point contributes at least an additional $\log\symWeightIndexMax$ factor. In contrast, the calculation time for \ac{ICWS} is only proportional to $\symHashSize$ in this case.

However, the BagMinHash algorithms are orders of magnitude faster for $\symInputSize \gg \symHashSize$, where their calculation times are approaching a linear function of $\symInputSize$. The gap to the calculation time of the \ac{ICWS} algorithm increases with $\symHashSize$, which indicates that the dominating term in the time complexity of BagMinHash does not depend on $\symHashSize$. 
The optimization described in \cref{sec:optimization} and implemented by BagMinHash~2 leads to a significant improvement of the calculation time for intermediate input sizes $\symInputSize$. The break-even point of BagMinHash~2 compared to \ac{ICWS} is around $\symInputSize = 100$.

As shown in \cref{fig:space}, we also investigated the space requirements in terms of Poisson process objects, that need to be simultaneously kept in memory. 
The maximum space required by BagMinHash~1 is dominated by the heap size needed for processing the very first element. Since the first element is faced with the largest $\symHashValueMax$ values, more points need to be generated and more Poisson process splittings are necessary than for later elements. This explains why the space requirements are essentially independent of $\symInputSize$. 
Our BagMinHash~2 implementation uses a max-heap as buffer, which allows efficient discarding of process objects while $\symHashValueMax$ is decreasing. This is the reason, why the space requirements are not very different to those of BagMinHash~1. Since the observed values are small enough and also the dependence on $\symHashSize$ seems to be at most quasilinear, the space requirements are not an issue for today's computers.
We did not investigate the space requirements of \ac{ICWS}. It only needs some vectors of size $\symHashSize$, which is negligible in practice.

The presented performance results refer to the single-precision weight discretization discussed in \cref{sec:discretization}. Since the discretization $\symDiscreteSet$ can be freely chosen, it can be optimized to achieve a certain precision, if all non-zero weights are from some known value range. For example, if an estimation error of $\varepsilon$ is acceptable for the Jaccard index, one could choose $\symDiscreteValue_\symWeightIndex = \symDiscreteValue_1 (1+\varepsilon)^{\symWeightIndex-1}$ with $1\leq\symWeightIndex\leq \symWeightIndexMax$ together with appropriate $\symDiscreteValue_1>0$ and $\symWeightIndexMax$ to cover the desired value range. In this way \eqref{equ:value_discretization} will be satisfied, while minimizing $\symWeightIndexMax$ and thus also computational costs.

\section{Conclusion}
We have presented BagMinHash, a new weighted minwise hashing algorithm for the calculation of hash signatures that can be used to estimate the generalized Jaccard similarity of weighted sets. Without being less accurate, BagMinHash is orders of magnitude faster for large input than the current state of the art. Therefore, we expect that our new approach will improve the performance of existing applications and also enable new use cases for which computational costs had been a limiting factor so far.

\bibliographystyle{ACM-Reference-Format}

\bibliography{bibliography.bib}

\end{document}